\documentclass[manuscript,nonacm, screen]{acmart}
\usepackage{amsmath,amsfonts}
\usepackage{array}
\usepackage{tabularx}
\usepackage{multirow}
\usepackage{enumitem}
\usepackage{placeins}
\usepackage{graphicx}
\usepackage{xcolor}
\usepackage{xspace}

\newcolumntype{L}[1]{>{\raggedright\arraybackslash}p{#1}}
\newcolumntype{C}[1]{>{\centering\arraybackslash}p{#1}}
\newcolumntype{Y}{>{\raggedright\arraybackslash}X}
\definecolor{mauve}{rgb}{0.25,0,0.52}
\definecolor{enEEG}{HTML}{D11E25}
\definecolor{enEye}{HTML}{2F7F92}
\definecolor{enEDA}{HTML}{A87C0A}
\newcommand{\qualquote}[2]{%
  \textit{\textcolor{mauve}{``#1''}}\nobreakspace%
  \textnormal{(#2)}%
}
\newcommand{\ent}{\texorpdfstring{\textup{\texttt{E\textsuperscript{3}Sense}}}{E3Sense}\xspace}

\AtBeginDocument{}

\setcopyright{none}
\copyrightyear{}
\acmDOI{}
\acmISBN{}
\acmConference{}{}{}
\acmBooktitle{}
\begin{document}

\title{\ent{}: Head-Confined Multimodal Sensing of Learner Engagement}



\author{Sidharth Anupkrishnan}
\authornote{Co-first authors. Both authors contributed equally to this research.}
\orcid{0000-0002-2441-5994}
\email{sanupkrishna@umass.edu}
\affiliation{%
  \institution{University of Massachusetts, Amherst}
  \city{Amherst}
  \state{Massachusetts}
  \country{USA}
}

\author{Itir Sayar}
\authornotemark[1]
\orcid{0009-0001-6133-9820}
\affiliation{%
  \institution{University of Massachusetts, Amherst}
  \city{Amherst}
  \state{Massachusetts}
  \country{USA}}
\email{isayar@umass.edu}

\author{Jeongah Lee}
\orcid{0000-0002-5714-5521}
\affiliation{%
  \institution{University of Massachusetts Amherst}
  \city{Amherst}
 \state{MA}
  \country{USA}
}
\email{jeongahlee@umass.edu}

\author{Sri Harsha Musunuri}
\orcid{0000-0003-1325-925X}
\affiliation{%
  \institution{Dolby Laboratories Inc.}
  \city{Sunnyvale}
  \state{California}
  \country{USA}}
\email{harsha.musu@gmail.com}

\author{Guan-Ming Su}
\orcid{0000-0002-3118-5904}
\affiliation{%
  \institution{Dolby Laboratories Inc.}
  \city{Sunnyvale}
  \state{California}
  \country{USA}}
\email{guanmingsu@ieee.org}

\author{Madeline Endres}
\orcid{0000-0002-4618-4939}
\affiliation{%
  \institution{University of Massachusetts, Amherst}
  \city{Amherst}
  \state{Massachusetts}
  \country{USA}}
\email{mendres@umass.edu}

\author{Ravi Karkar}
\authornote{Co-corresponding authors.}
\orcid{0000-0003-1467-4439}%
\affiliation{%
 \institution{University of Massachusetts Amherst}
 \city{Amherst}
 \state{MA}
 \country{USA}}
 \email{rkarkar@umass.edu}
 
\author{Phuc Nguyen}
\authornotemark[2]
\orcid{0000-0001-8078-4463}
\affiliation{%
  \institution{University of Massachusetts, Amherst}
  \city{Amherst}
  \state{Massachusetts}
  \country{USA}}
\email{vp.nguyen@cs.umass.edu}

\renewcommand{\shortauthors}{Anupkrishnan and Sayar et al.}

\begin{abstract}
 Engagement-aware learning systems could provide hints or adjust pacing when learners struggle. Prior engagement sensing work distributes sensors across or outside the body rather than consolidating them at one site, or reduces engagement to shared affect or a single dimension (from behavioral, emotional, and cognitive engagement). We introduce \ent, a head-worn platform that co-locates electroencephalography, eye tracking, and electrodermal activity to personalize engagement measurement. During a lab study we collected 450 ratings of engagement levels on a five-level ordinal scale while participants watched educational videos. For fifteen held-out participants, \ent  achieved a within-one-level prediction score of $75.0\%$, compared with $63.0\%$ for always predicting
the most common rating. In an exploratory analysis of 18 participants from the same study who defined engagement, conditioning on learners' definitions raised the same measure by $6.9$ points, from $64.6\%$ to $71.5\%$. Our work provides a proof-of-concept of a head-site, personalized multimodal sensing of engagement for adaptive educational interfaces.
\end{abstract}



\begin{teaserfigure}
 \centering
  \includegraphics[width=\textwidth]{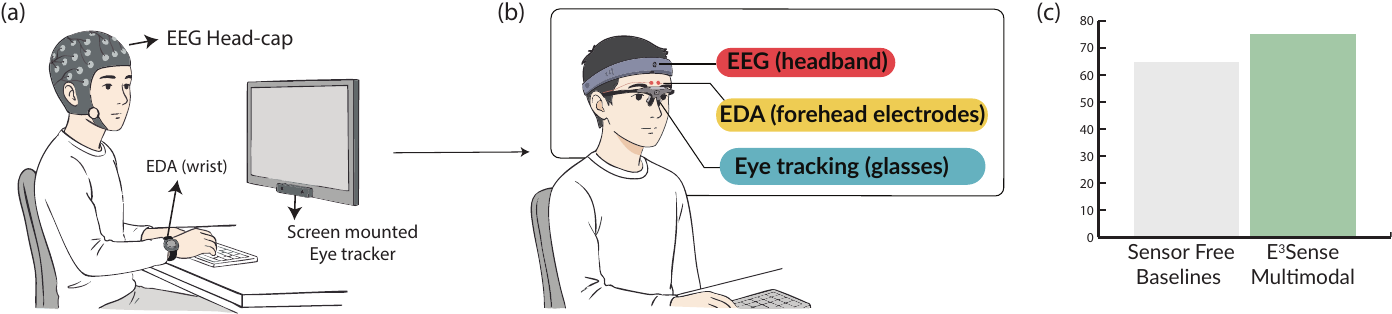}
\caption{\textbf{\ent confines engagement sensing to the head.}
(a)~Conventional multimodal sensing spreads electroencephalography (EEG), wrist electrodermal activity (EDA), and a screen-mounted
eye tracker across body and desk. (b)~\ent places all three on the head: an EEG
headband, forehead EDA electrodes, and eye-tracking glasses.
(c)~Across participant-grouped folds evaluating fifteen forehead-sensing learners,
\ent achieves a within-one-level prediction score of $75.0\%$, compared with
$63.0\%$ for a sensor-free reference that always predicts the most common
training-fold rating. The score gives equal weight to each represented rating
level. A separate exploratory analysis examines whether learner-definition
codes improve prediction relative to matched score calibration
(Section~\ref{sec:rq2_results}).} 
\Description{Overview of the engagement sensing system, showing conventional and head-based sensor placements, and a summary of prediction results.}
\label{fig:teaser}
\end{teaserfigure}

\maketitle
\section{Introduction}
\label{sec:intro}

Videos are widely used in online learning~\cite{guo2014video}. When learners
watch prerecorded educational videos independently, an instructor may not be
available to recognize confusion and adjust the pace. Engagement is associated
with academic achievement~\cite{lei2018relationship}, but different measures
provide different views of the learning experience. Quiz responses, clickstream
traces, pauses, and replays capture aspects of learning
behavior~\cite{guo2014video}, while questionnaires capture learners' reported
experiences~\cite{fredricks2012measurement,henrie2015measuring}. Retrospective
reports depend on recall, and prompt pop-ups during viewing can interrupt
the experience being measured~\cite{gao2023critiquing,krosnick1999survey}.
Physiological and behavioral sensing offers an additional source of
continuously recorded signals. Such signals could inform systems that offer
hints, adjust pacing, or suggest breaks, provided that the estimates are
meaningful and available at an appropriate
timescale~\cite{pope1995biocybernetic,giannakos2019multimodal,bustos2022wearables}.

However, continuous multimodal sensing can introduce substantial instrumentation and setup requirements.
Prior multimodal sensing configurations
distribute their sensors across the learner and the environment: a multi-electrode
EEG cap on the scalp, an electrodermal sensor on the wrist or fingers, and an
eye tracker mounted beneath the display
(Figure~\ref{fig:teaser}a)~\cite{ji2025senseseek,dwivedi2024effecti}. While such an
arrangement suits laboratory experiments, maintaining and calibrating distributed sensors presents a barrier for deploying such systems in practice (e.g., watching a lecture recording in a
dormitory, or attending a university course).
Confining the sensor set to familiar head-worn form factors could reduce this deployment burden.
This raises a feasibility question: can signals collected entirely at the head predict self-reported engagement in previously unseen learners? Implementing it with commercially available devices would also make the approach more accessible for replication and further development. We therefore investigate the technical feasibility of head-confined multimodal engagement sensing using off-the-shelf devices, providing a foundation for subsequent studies of usability and everyday deployment.

Complementing this investigation of sensor placement, we examine whether learners’ own definitions of engagement can help contextualize sensor-based predictions. Educational psychology characterizes engagement through behavioral, emotional, and cognitive dimensions~\cite{fredricks2012measurement,gao2023critiquing}. When providing an overall engagement rating, learners may emphasize different aspects of that experience, such as sustained attention, understanding, or emotional involvement. We explore whether these self-described criteria provide useful context for mapping physiological and behavioral signals to learners’ engagement ratings.

In this paper we present \textbf{\ent} (Engagement via EEG, eye tracking, and
EDA), a head-worn multimodal sensing platform. \ent co-locates three
devices on the head: a dry-electrode EEG headband, eye-tracking
glasses, and an EDA sensor with dry forehead electrodes. It aligns their three
streams in software and predicts a learner's segment-level engagement rating from a
single fused representation of all three
(Section~\ref{sec:system_overview}). All sensing sites lie on the head, and the EEG and EDA electrodes are dry.
This layout could inform a future integrated device. To evaluate
\ent, we ask two research questions.

\begin{itemize}[leftmargin=*]
\item \textbf{RQ1:} How accurately does \ent predict engagement for unseen learners, and how does performance
vary across conventional model families?
\item \textbf{RQ2:} How do learners define engagement, and what is observed when those definitions are supplied as context to an exploratory calibration model? 
\end{itemize}
Two analyses of the same study answer these questions under separate evaluation protocols (Section~\ref{sec:evaluation}).  
RQ1 is the paper's central protocol: five
participant-grouped folds train across both EDA collection sites and score the
fifteen held-out forehead participants so every rating
scored comes from a learner no model was fitted to. RQ2 is a smaller,
exploratory evaluation over the eighteen participants who gave definitions of engagement in their own words, under its own six-fold protocol. The two studies draw on
different cohorts and different procedures, so we report them separately and do
not compare their numbers with each other.

\ent achieves 75.0\% balanced one-off accuracy, which gives equal weight to each engagement class and counts predictions within one level of the learner’s rating as correct. This exceeds the sensor-free baseline of 63.0\% by 12.0 percentage points (Section~\ref{sec:rq1_results}). In the complementary analysis, learners described engagement through cognitive, behavioral, and emotional criteria. Incorporating these criteria into an exploratory calibration model increased balanced one-off accuracy from 64.6\% to 71.5\% relative to a matched model without this context (Section~\ref{sec:rq2_results}). Together, these findings support the technical feasibility of head-confined multimodal engagement sensing and motivate further investigation of learner definitions as context for personalized engagement prediction.

\noindent Our contributions are:
\begin{enumerate}[leftmargin=*]
\item \textbf{\ent, a head-worn multimodal sensing platform, and a
participant-independent evaluation of head-confined engagement prediction.}
\ent co-locates dry-electrode EEG, eye tracking, and dry forehead EDA on the
head and fuses them into a fixed 166-dimensional per-segment representation. We
evaluate that representation with five classical model families and sensor-free
baselines on held-out forehead participants. AdaBoost attains $75.0\%\pm7.0$
balanced 1-off accuracy and $1.043\pm0.158$ macro-MAE, a fold mean $12.0$
percentage points above the mode baseline, and LightGBM holds the highest
binary macro-F1 point estimate at $58.9\%\pm7.9$ (\S\ref{sec:rq1_results}).

\item \textbf{Learner definitions of engagement, coded qualitatively, with an
exploratory predictive probe.}
Participants' coded accounts span cognitive, behavioral, and emotional
definitions. Supplying those codes as context to the second-stage ordinal model
attains $71.5\%$ balanced 1-off accuracy against $64.6\%$ for a matched
control, an observed $+6.9$-percentage-point difference alongside lower
macro-MAE. Within this eighteen-participant cohort, what a learner counts as
engagement carries information about their ratings that the sensor-derived
scores alone do not (\S\ref{sec:rq2_results}).
\end{enumerate}

\section{Background and Related Work}

Prior work relevant to \ent spans three linked questions: how engagement is conceptualized, how engagement labels are elicited and evaluated, and how physiological sensing modalities are configured. We review these threads to motivate two gaps: (i) whether multimodal engagement sensing can be confined to the head while retaining participant-independent predictive utility, and (ii) whether learner-specific definitions can provide context for the target being predicted.

\subsection{Background on Engagement}
\label{sec:what_is_engagement}

Engagement is a complex multi-dimensional construct, where its specific components and associated metrics are domain-dependent (see Doherty et al. for a review of engagement theories and measurement~\cite{doherty2018engagement}).  
One framework commonly used in educational psychology is the \textit{tripartite framework}~\cite{fredricks2004school}, which identifies three interconnected aspects of engagement: behavioral participation and attention,
emotional interest and enjoyment, and cognitive effort and depth of
processing~\cite{fredricks2004school}. These dimensions are related but
dissociable~\cite{benEliyahu2018multidimensionality,booth2023engagement} and have
associations with academic achievement~\cite{lei2018relationship}. We use this
framework to motivate \ent and to discuss learners'
accounts, while targeting a single overall self-reported engagement rating as
our prediction outcome. Automatic engagement
detection has also been studied through face- and gaze-based recognition
during learning~\cite{de2019engaged, whitehill2014faces, karimah2022automatic}, affect-in-the-wild
competitions~\cite{dhall2018emotiw}, and reviews of multimodal
learning-experience data~\cite{giannakos2019multimodal} and wearable sensing
for engagement specifically~\cite{bustos2022wearables}. We build on this
literature's physiological and behavioral signals.

\subsection{Engagement Labels and Evaluation}

Engagement is commonly measured through questionnaires, post-task surveys, and
experience sampling~\cite{fredricks2012measurement}, including standardized
affect instruments such as the Self-Assessment
Manikin~\cite{bradley1994measuring}. The temporal unit matters.
A session-level response summarizes more activity than a segment rating, while
retrospective reports cannot directly resolve moment-to-moment
changes~\cite{gao2023critiquing}. Physiological and behavioral signals offer
continuous correlates, but relating them to a rating depends on what the
rating denotes~\cite{gao2020n,gao2022understanding}. We collect learners' own
accounts to contextualize that target and examine their predictive use.

Engagement labels are often unevenly distributed~\cite{gupta2016daisee,
singh2023engagenet}. Frequency-weighted performance and class-balanced ordinal
performance answer different questions~\cite{baccianella2009evaluation}.
We report balanced 1-off accuracy and macro-MAE alongside sensor-free baselines,
with binary macro-F1 as a secondary sensitivity analysis
(Section~\ref{sec:metrics}). This makes the contribution of the label
distribution explicit without treating near-correct ordinal predictions as
exact five-class decisions.

Participant-disjoint evaluation is also consequential. Physiological variation
between people can make performance on familiar participants different from
performance on unseen ones~\cite{saeb2017approximate}. Engagement studies have
used both participant-disjoint protocols~\cite{diLascio2018engagement,
gao2020n,dwivedi2024effecti,monkaresi2016automated} and segment-level
splits~\cite{xiao2023multimodal}. We hold participants out whole in our outer
evaluation, the stricter protocol and the one that speaks to how the model
generalizes to new learners. RQ2 uses their exit-interview definition codes as context.

\subsection{Multimodal and Head-Worn Sensing}
\label{sec:modeling_engagement}

EEG, eye tracking, and EDA provide candidate correlates of engagement. EEG band
power is associated with attention and cognitive processing~\cite{klimesch1999eeg,
pope1995biocybernetic}, including in classroom
settings~\cite{ramirez2021eeg}; gaze describes visual orientation and pupil dynamics
reflect several influences, including cognitive load~\cite{rayner1998eye,
beatty1982task,kahneman1973attention}. EDA reflects sympathetic
activation~\cite{boucsein2012electrodermal}. Multimodal physiological datasets
and frameworks combine EEG with other channels for affect
recognition~\cite{koelstra2011deap,zheng2019emotionmeter}, but
none provides a unique reading of one engagement dimension. We therefore evaluate
the complete fused \ent representation against the same overall rating across five
classical model families.

Placement distinguishes existing configurations. SenseSeek combines head-worn
EEG, wrist EDA, and screen-mounted eye tracking for information
seeking~\cite{ji2025senseseek}. Effecti-Net combines head-worn EEG and eye
tracking with finger EDA and PPG to predict perceived content
effectiveness~\cite{dwivedi2024effecti}. Long et al.\ combine EEG and eye
tracking to detect internal versus external attention in
VR~\cite{long2024multimodal}, and Brishtel et al.\ combine eye tracking and
EDA to detect mind wandering during reading~\cite{brishtel2020mind}. Head-worn
platforms such as Galea and
HMDBioPad demonstrate co-located instrumentation~\cite{bernal2022galea,
wan2022hmdbiopad}, while scalp EEG with forehead EOG has been evaluated for
vigilance~\cite{zheng2017multimodal}. These motivate examining what EEG, eye
tracking, and forehead EDA together predict about engagement during learning.

Table~\ref{tab:relatedwork_comparison} summarizes placement, targets, and
labeling practices in the studies considered here. Of these, Knierim et
al.~\cite{knierim2025advancing} and Kosmyna et al.~\cite{attentivu} confine every sensing site to the head but each uses
a single sensing modality (EEG), while Galea~\cite{bernal2022galea} and HMDBioPad~\cite{wan2022hmdbiopad} are head-confined
multimodal platforms not evaluated for engagement prediction. \ent is, to our
knowledge, the first configuration in this comparison set to combine three
head-confined modalities, evaluate them against segment-level engagement
ratings, and pair the sensing results with learners' own definitions of
engagement. We compare five classical model families on the same fused,
handcrafted-feature representation, keeping the sensor input fixed throughout
RQ1.

\begin{table*}[htbp]
\centering
\footnotesize
\renewcommand{\arraystretch}{1.15}
\setlength{\tabcolsep}{4pt}

\caption{Comparison of Student Engagement Sensing and Labeling Practices in Prior Work.}
\Description{An eight-column table comparing ten prior sensing
studies with \ent. The columns record, for each work, its data sources and sensor
placement, whether every sensor is confined to the head, which engagement
dimensions it targets, session duration, the feedback instrument used for
labels, the temporal unit each label covers, and the granularity of the
engagement scale. Four prior works confine every sensor to the head: two use a
single modality with a cognitive target coarser than five levels, and two are
multimodal head-worn platforms that do not evaluate engagement prediction. The
\ent row combines head-confined multimodal sensing with segment ratings and
learner-definition interviews.}
\label{tab:relatedwork_comparison}

\begin{tabular}{@{}
L{1.85cm}
L{2.40cm}
C{1.20cm}
L{1.50cm}
C{1.20cm}
L{2.30cm}
C{1.35cm}
C{1.35cm}
@{}}

\toprule
\textbf{Work} &
\textbf{Data Sources} &
\textbf{Head-Confined} &
\textbf{Eng. Framing} &
\textbf{Duration} &
\textbf{Feedback Type} &
\textbf{Temporal Resolution} &
\textbf{Eng. Granularity} \\
\midrule

Di Lascio et al.~\cite{diLascio2018engagement} &
EDA (wrist) &
No &
Emotional &
45 min &
Std. Q. &
Lecture &
Binary \\

Monkaresi et al.~\cite{monkaresi2016automated} &
Video &
No &
Behavioral &
60 min &
Self-report &
\textbf{2\,min / 10\,s} &
Binary \\

Dwivedi et al.~\cite{dwivedi2024effecti} &
EDA/PPG-f; eye/EEG-h; cam &
No &
--- &
9 min &
Ratings &
\textbf{Segment} &
Likert-3 \\

Knierim et al.~\cite{knierim2025advancing} &
EEG (head) &
\textbf{Yes} &
Cognitive &
45 min &
Std. Q. &
Task &
Ternary \\

Ji et al.~\cite{ji2025senseseek} &
EDA-w; eye (screen); EEG-h &
No &
--- &
--- &
--- &
--- &
--- \\

Kosmyna et al.~\cite{attentivu} &
EEG-h; haptic scarf (neck) &
\textbf{Yes}\textsuperscript{$\dagger$} &
Cognitive &
105 min &
Self-report + quiz &
\textbf{Slide} &
Binary \\

Gao et al.~\cite{gao2020n} &
EDA/HRV-w; env. sensors &
No &
\textbf{Behav.\slash emot.\slash cog.} &
40--80 min &
Std. Q. &
Class &
Likert-5 \\

Xiao et al.~\cite{xiao2023multimodal} &
EEG-h; eye (screen); cam &
No &
\textbf{Behav.\slash emot.\slash cog.} &
40 min &
Std. Q. &
Session &
Likert-5 \\

Bernal et al.~\cite{bernal2022galea} &
Physiological suite (head, HMD) &
\textbf{Yes} &
None &
--- &
--- &
--- &
--- \\

Wan et al.~\cite{wan2022hmdbiopad} &
Bio-signal pads (head, HMD) &
\textbf{Yes} &
None &
--- &
--- &
--- &
--- \\

\midrule

\textbf{\ent (this paper)} &
\textbf{EEG/eye/EDA (head)} &
\textbf{Yes} &
Overall rating; three-dimension definitions &
36 min &
\textbf{Self-report + interview} &
\textbf{Segment} &
Likert-5 \\

\bottomrule
\end{tabular}

\parbox{\textwidth}{\scriptsize
\textit{Abbreviations:}
w = wrist; f = finger; h = head; cam = webcam; env. = environmental;
cog. = cognitive; behav. = behavioral; emot. = emotional;
std. Q. = standardized questionnaire.
The temporal-resolution column reports the unit each engagement label covers.
\textsuperscript{$\dagger$}AttentivU's scarf delivers haptic feedback and carries
no sensor, so every sensing site is on the head.
Dashes mark properties the work does not report; where a system targets a
construct other than engagement, as SenseSeek and Effecti-Net do, the
engagement-specific columns do not apply to it.}

\end{table*}

\section{The \ent Sensing Configuration}
\label{sec:system_overview}

\subsection{Design Goals}

The configuration was designed to co-locate neural, ocular, and electrodermal
sensing on the head and evaluate the complete configuration in a controlled
video-learning study. Dry electrodes and head-worn devices make this placement
practical to deploy.

\subsection{Architecture}
\label{sec:en3_platform}

\textbf{\ent} co-locates three
devices on the head and aligns their streams in software, establishing a
sensor layout that a future integrated device can adopt directly.
Figure~\ref{fig:teaser} diagrams the sensor placement;
Section~\ref{sec:algorithms} covers feature extraction and prediction.


\subsubsection{Sensing Layer}

Three consumer-grade devices supply the three streams.

\begin{itemize} [leftmargin=*]
    \item \textbf{EEG headband.}
     The Frenz BrainBand~\cite{frenzbrainband} delivers four dry-electrode
    channels over Bluetooth at 125\,Hz: two frontal (LF, RF) and two temporal
    over the ears. Only the frontal pair enters the analysis, for its
    established sensitivity to attention regulation and cognitive
    control~\cite{pope1995biocybernetic}. We use dry electrodes for faster,
    gel-free setup~\cite{casson2019wearable}; band-power features of the kind
    we extract remain reliable with dry electrodes~\cite{kleeva2024dryeeg}.
    
    \item \textbf{Eye tracker.}
    The Pupil Labs Neon~\cite{pupillabsneon} captures infrared eye video at
    200\,Hz and processes it on-device into a twenty-channel 3D eye state: pupil
    diameter, eyeball-center and optical-axis coordinates for both eyes, and
    eyelid angles and apertures. We retain all twenty.

    \item \textbf{EDA sensor.}
    The EmotiBit~\cite{montgomery2024validating} measures EDA at 15\,Hz. To move
    the measurement to the head we replaced its 
    electrodes with two dry silver/silver-chloride electrodes on the forehead just above
    the eyebrows, the eye-tracker frame fixing their separation at
    $\approx$4\,cm so spacing is identical across participants. Prior comparisons characterize forehead EDA as relatively resistant to
    motion artifacts but less responsive than palmar recordings, motivating a
    feasibility evaluation rather than an equivalence claim~\cite{Hossain2022Comparison}.

\end{itemize}

The three devices clock independently, so we align the streams on device-native
timestamps and periodic clock anchors collected during recording, making
features from one window traceable to the same stretch of viewing in all three.

\subsubsection{Preprocessing Layer}
\label{sec:preprocessing_layer}

Each stream reaches the feature extractor filtered for its sensor's characteristic
  artifacts, with the frequency content its features depend on retained. For EEG, the
  headband delivers the frontal channels band-limited from 0.5 to 60\,Hz, a passband
  chosen on standard filter-design grounds~\cite{widmann2015digital}: it removes the DC drift that dry contacts introduce while retaining the canonical bands ($\delta$ 1 to 4\,Hz, $\theta$ 4 to 8, $\alpha$ 8 to 13, $\beta$ 13 to 30, and $\gamma$ 30 to
  45\,Hz), from whose token spectrum we compute band power. For eye tracking, we use
  the twenty eye-state channels directly as the device exports
  them~\cite{pupillabsneon}. The eye features in Table~\ref{tab:features} are
  therefore statistics of these per-sample channels plus a few derived pupil and blink quantities; where a threshold is needed, we take it from the token's own median. For EDA, a fourth-order Butterworth lowpass at 0.5\,Hz attenuates higher-frequency components while retaining the tonic conductance level and the slow phasic envelope~\cite{boucsein2012electrodermal}; we decompose the filtered signal into these tonic and phasic components, both of which carry information about
  sympathetic arousal~\cite{matton2023contrastive}.

\subsubsection{Tokenization}
Two units of time recur here and are distinct. A \textit{segment} is the rated
unit of Section~\ref{sec:study_overview}. An \textit{analysis window} is the
fixed 120\,s interval from the start of each segment, from which we compute that
segment's features, so every observation carries an equal span of signal
regardless of video length. Within that window we use twelve 10\,s EEG tokens, six 20\,s eye-tracking
tokens, and three 40\,s EDA tokens. EEG band power uses Welch estimates over
4\,s sub-segments. Eye tokens summarize the device-exported channels and derived
pupil and blink quantities. EDA tokens use a 10\,s centered rolling median for
the tonic component. Features are mean-pooled across tokens within each
modality, giving one vector per segment's analysis window.

\section{Study Overview}
\label{sec:study_overview}
\begin{figure*}[htbp]
\centering
\includegraphics[width=\textwidth]{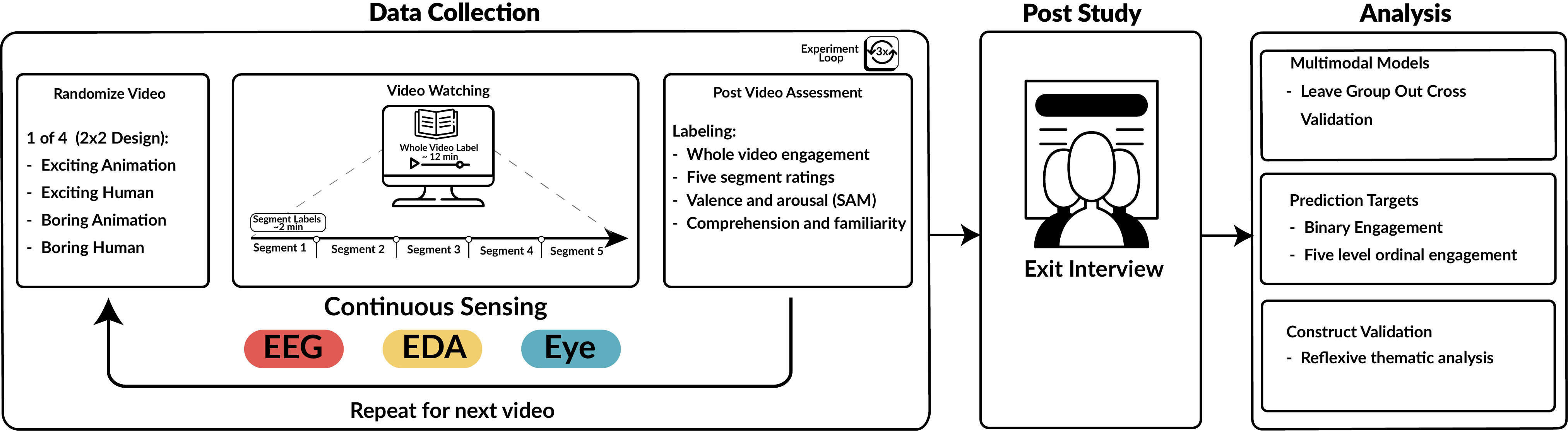}
\caption{ Collection and analysis procedure. Participants watched three of
four videos in randomized order under continuous EEG, eye, and EDA sensing,
completing a post-video assessment that includes five segment ratings and
additional video-level measures; the loop repeats three times. The leading
120\,s of each rated segment supplies the modeled features. EDA was recorded
at the wrist for the first fifteen participants and the forehead for the
next fifteen. 24 participants additionally completed an exit interview out of which 18 gave definitions of engagement in their own words. Five classical model families
using the fused \ent representation are then evaluated under participant-grouped
cross-validation, predicting binary and five-level ordinal engagement.}
\Description{Three panels: Data Collection, Post Study, and Analysis. Data
Collection shows video randomization across a two-by-two content design,
video watching under continuous EEG, EDA, and eye sensing, and a post-video
assessment recording five segment ratings and additional measures, looped
three times for 30 participants, 3 videos, and 5 segments per video, yielding
450 labeled segments. Analysis shows five classical model families using the
fused \ent representation evaluated under participant-grouped cross-validation,
predicting binary and five-level ordinal engagement.}
\label{fig:study_procedure}
\end{figure*}

The study addressed two limitations of common engagement-measurement approaches. First, session or video level ratings can obscure changes in engagement within a video. We therefore asked participants to retrospectively rate five segments after each video. Second, standardized questionnaires define engagement in advance and may not capture how individual learners interpret the construct. We therefore conducted semi-structured exit interviews in which participants described what engagement meant to them and how they formed their ratings. Both components were completed within a single study session (Figure~\ref{fig:study_procedure}). A single session served both questions. We first describe the collection procedure common to them, then the sensor placement that RQ1 turns to, and finally the interviews that RQ2 draws on. 


\subsection{Data Collection Common to Both Questions}
\label{sec:procedure}

\subsubsection{Participants and Set-up}
\label{sec:participants}
We recruited thirty participants ($N=30$) from a university campus. Ages ranged
from 19 to 43 years (median 26, $SD=5.86$); 11 were female and 19 male. We
required English fluency, vision normal or corrected by contact lenses, and no
history of neurological conditions. Eyeglasses occlude the eye tracker's
infrared cameras, and neurological history can introduce confounds in
EEG signals. Each received a \$20 gift card for about 60 minutes. The
institution's Institutional Review Board (IRB) approved the protocol. On arrival participants watched a
device-fitting demonstration, gave informed consent, then sat at a 27-inch
monitor and put on the recording devices, whose signal quality we verified before viewing began.

\subsubsection{Video Stimuli and Assignment}
\label{sec:video_stimuli}
Four publicly available educational videos populate a $2\times2$ design crossing
content type (Exciting/Boring) with presentation format (Animation/Human), one
per cell: Exciting Animation (12:08, 728\,s), Exciting Human (12:54, 774\,s),
Boring Animation (12:04, 724\,s), and Boring Human (11:41, 701\,s), covering
human anatomy, gene modification, popular science, and linear algebra. Content
type was assigned a priori, before any data collection: four researchers viewed
the candidates and settled by discussion which were exciting and which boring.

The four videos vary content and presentation. The four cells were used to diversify the viewing stimuli rather than to support factorial inference; with one video per cell, we do not estimate independent effects of content type or presentation format.
 The roughly 12-minute
length sits well above the under-six-minute chunks MOOC engagement research
recommends~\cite{guo2014video} --- one reason we rate per segment. Each
participant watched three of the four, selection and order randomized in
Qualtrics, giving 90 video observations: Exciting Animation ($N=21$), Exciting
Human ($N=25$), Boring Animation ($N=21$), and Boring Human ($N=23$).

\subsubsection{Engagement Labeling}
Participants rated engagement retrospectively, immediately after each video
rather than continuously during viewing, following standard
practice~\cite{fredricks2012measurement}; this keeps the viewing itself
uninterrupted and more natural, without introducing confounding factors that would affect the state we want
to capture. They first gave an overall engagement rating on a five-point
Likert scale (Extremely Disengaged to Extremely Engaged). This granularity was
chosen to distinguish intensities without imposing too much load across
repeated ratings~\cite{krosnick1999survey}. Then they rated engagement for five
equal segments: for a video of duration $T$, segment $k$ spans
 $[\,kT/5,\;(k+1)T/5\,]$ for $k=0,\dots,4$, some 140 to 155\,s. The cohort
supplies 450 segment-level observations, the prediction target throughout,
because the whole-video rating averages over exactly the fluctuations an engagement
sensing system wants to detect.

We code the five categories as 0 (Extremely Disengaged), 1 (Disengaged),
2 (Neutral), 3 (Engaged), and 4 (Extremely Engaged). The label distribution
skews toward engagement: 37, 57, 73, 169, and 114 observations from
\textit{Extremely Disengaged} to \textit{Extremely Engaged}, so 63\% of the
ratings sit at the two engaged levels and 21\% at the two disengaged ones. We
therefore report the main prediction metrics in Section~\ref{sec:metrics} with
class-balanced averaging, so performance on the rarer disengaged states is not
masked by the majority classes.

\subsection{RQ1: Head-Confined \ent Evaluation}
\label{sec:rq1_study}
Confining the sensor set to the head moves the EDA electrodes from the wrist to
the forehead (Section~\ref{sec:system_overview}). Fifteen participants wore wrist EDA, the conventional site~\cite{gao2020n, diLascio2018engagement}, and the remaining
fifteen wore forehead EDA. No participant wore both, so we report no
wrist--forehead contrast. Models train on both sites because head confinement
constrains where sensors sit during use, not where the training data came from.
Every learned RQ1 row uses the same 166-dimensional \ent representation. Every
model-family result in Table~\ref{tab:main_results} is scored on
the fifteen forehead participants alone (Section~\ref{sec:cross_validation}). The definition-context analysis scores
all eighteen coded participants, the seven wrist-EDA participants included,
since its cohort follows the interviews, not sensor site
(Section~\ref{sec:definition_context_methods}).

\subsection{RQ2: Collecting Learner Definitions} 

\label{sec:interviews}
Six participants out of 30 left the session before the interview could be
finalized, due to unforeseeable conflicts or time constraints. The remaining participants completed semi-structured interviews and a subset of 18 participants defined engagement in their own words. In the interviews we asked how they had decided their ratings, what
raised or suppressed their engagement, and how they would describe engagement in
their own words. These accounts supply the learner-defined descriptions of
engagement that RQ2 draws on, as well as context for interpreting the
quantitative ratings.

\section{Algorithms}
\label{sec:algorithms}

\ent extracts handcrafted features from each stream and concatenates them into a
fixed representation for classical prediction models. Below we describe: (i) the
preprocessing and feature-extraction pipeline shared by both questions, (ii) the
fused \ent representation and five model families evaluated in RQ1, and (iii) the
second stage through which a learner's own definition of engagement enters the
RQ2 model.

\subsection{Common to Both Questions}
\label{sec:algorithms_shared}

\subsubsection{Modeling Overview}
\label{sec:architecture_overview}

\begin{figure*}[htbp]
  \centering
  \resizebox{\textwidth}{!}{%
  \includegraphics{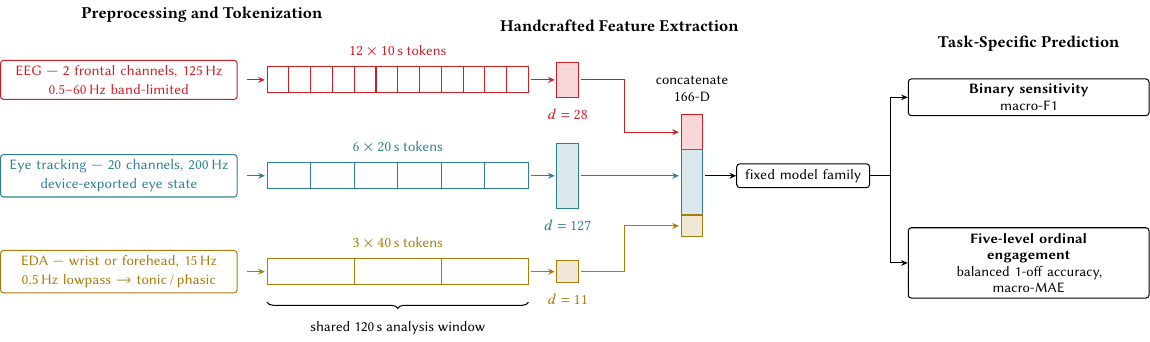}
  }%
  \caption{The \ent processing pipeline. We preprocess each stream, tokenize it,
  compute handcrafted features from every token, mean-pool them into one vector
  per modality and window, and pass the result to classical models trained
  separately for binary and five-level ordinal engagement. RQ1 evaluates five
  model families as separate rows, keeping the 166-dimensional \ent input fixed
  (Section~\ref{sec:training_procedure}).}
  \Description{Block diagram of the processing pipeline, drawn as three parallel
horizontal lanes, one per modality, colored consistently with the rest of the
paper. Each lane runs from its raw stream through preprocessing to a row of
fixed-length tokens, twelve for EEG, six for eye tracking, and three for EDA,
all spanning the same shared analysis window. Handcrafted features are computed
per token and mean-pooled into one vector per modality. The three lanes then
converge into a single concatenated vector, which passes to a classical model
before forking to two prediction heads, binary engagement and five-level
ordinal engagement.}
  \label{fig:pipeline}
\end{figure*}

The 120\,s analysis window of Section~\ref{sec:en3_platform} is set by the
40\,s EDA token. The window must hold a whole number of them and still fit
inside the shortest segment of 140\,s, and 120\,s also divides evenly into the
10 and 20\,s tokens. Within each analysis window the pipeline runs in three stages
(Figure~\ref{fig:pipeline}): (i) preprocessing and tokenization
(Section~\ref{sec:preprocessing_layer}), (ii) feature extraction and mean-pooling,
and (iii) task-specific prediction. Every RQ1 model family receives the same
166-dimensional fused representation and is evaluated under the same protocol.
RQ2 adds a second stage over the pipeline's outputs rather than altering the pipeline itself
(Section~\ref{sec:definition_context_model}).

\subsubsection{Handcrafted Feature Extraction}
\label{sec:handcrafted_features}
We extracted the handcrafted features summarized in Table~\ref{tab:features}
independently from each sensor-specific token and mean-pooled corresponding
features across the 120\,s analysis window. The 28-dimensional EEG representation
combines time-domain statistics, Hjorth parameters, frequency-band power estimated
using Welch's method, and spectral entropy~\cite{welch1967fft,long2024multimodal}.
The 127-dimensional eye-tracking representation summarizes the twenty
device-exported eye-state channels together with derived blink, gaze-dispersion,
and pupil-dynamics measures used in related attention-sensing
work~\cite{long2024multimodal,brishtel2020mind}. The 11-dimensional EDA
representation characterizes the raw, tonic, and phasic signals using statistical,
response-count, amplitude, area, and trend features previously applied in classroom
and engagement sensing~\cite{gao2020n}. Concatenating the three representations
yields the fixed 166-dimensional \ent vector supplied to every learned RQ1 model.

\begin{table*}[htbp]
\centering
\small
\caption{Handcrafted features extracted from each modality. We compute features
within every temporal token and mean-pool them across the analysis window, giving
one vector per modality and window; $d$ is the resulting dimensionality.}
\Description{Table listing the handcrafted features computed for each of the three
modalities, one block per modality, with the feature families named in each row
and the resulting per-modality dimensionality given at the block level. The three
blocks differ substantially in width, the eye-tracking block being much the
largest.}
  \label{tab:features}

\begin{tabular}{@{}L{2.0cm}c L{11.5cm}@{}}
\toprule
\textbf{Modality} & \textbf{$d$} & \textbf{Features} \\
\midrule

EEG & 28 &
Per frontal channel (LF, RF), 14 each: mean, standard deviation, root-mean-square
and peak-to-peak amplitude, zero-crossing rate, Hjorth activity, mobility and
complexity, log band power in the canonical bands ($\delta$ 1 to 4\,Hz,
$\theta$ 4 to 8\,Hz, $\alpha$ 8 to 13\,Hz, $\beta$ 13 to 30\,Hz,
$\gamma$ 30 to 45\,Hz),
and spectral entropy of the normalized power spectrum. \\
\addlinespace

Eye tracking & 127 &
Six statistics (mean, standard deviation, peak-to-peak range, mean absolute
successive difference, and the 25th and 75th percentiles) for each of twenty
per-sample channels: pupil diameter, eyeball-center and optical-axis coordinates
($x$, $y$, $z$), eyelid angles (top and bottom) and eyelid aperture, left and
right (120 dimensions). Derived, left eye: blink fraction, the share of samples
whose eyelid aperture falls below half its token median; gaze dispersion, the
summed standard deviation of the optical-axis $x$ and $y$ components; mean pupil
dilation and constriction rate; the fraction of invalid pupil samples; and the
mean and standard deviation of pupil diameter (7 dimensions). \\
\addlinespace

EDA & 11 &
Mean and standard deviation of the raw signal and of its tonic and phasic
components; skin conductance response (SCR) count; mean and maximum SCR
amplitude; area under the positive phasic curve; and the linear trend across the
window. \\

\bottomrule
\end{tabular}
\end{table*}

\subsubsection{Data Quality Check}
Following established EDA quality-assessment practices
\citep{kleckner2018simple,babaei2021critique}, we excluded four
participant-session EDA recordings---two wrist and two forehead---because the
recording was missing owing to battery failure or contained fewer than 30 finite
samples, our prespecified minimum-data threshold. The four sessions account for
20 of the 450 study segments. Two affected sessions belong to forehead
participants, producing 10 rows among the 225 RQ1 test segments ($4.4\%$) without
measured EDA. All eleven EDA features are missing in affected rows, so we impute
each feature from its median among the training participants in that fold. EEG
and eye-tracking features are complete for every segment. Section~\ref{sec:rq1_results}
reports a sensitivity analysis that excludes affected sessions from model fitting
and evaluation.

\subsection{RQ1: Models for the Head-Confined \ent Representation}
\label{sec:rq1_models}

\subsubsection{\ent Feature Representation}
\label{sec:en3_feature_representation}

For multimodal prediction we use early feature-level
fusion~\cite{baltrusaitis2019multimodal}, concatenating the mean-pooled EEG,
eye-tracking, and EDA feature vectors for the same analysis window:

\begin{equation}
\mathbf{x}_{\mathrm{multi}}
=
\left[
\mathbf{x}_{\mathrm{EEG}};
\mathbf{x}_{\mathrm{Eye}};
\mathbf{x}_{\mathrm{EDA}}
\right],
\end{equation}

giving a 166-dimensional representation from 28 EEG, 127 eye-tracking, and 11
EDA features. This complete vector is the fixed input for every learned RQ1 row.

\subsubsection{Model Families}

We evaluate a linear model, LightGBM~\cite{lightgbm2017},
XGBoost~\cite{xgboost2016}, AdaBoost~\cite{adaboost1997}, and Random
Forest~\cite{randomforest2001}. These families are commonly used with
handcrafted physiological features (Section~\ref{sec:modeling_engagement}), and
the four tree ensembles suit a few hundred labeled observations with mixed scales
and correlated predictors. Each family appears in a classifier or regressor
variant according to the target, with fixed within-family hyperparameters. Each
family is reported separately and remains fixed across the five outer folds.

\subsubsection{Sensor-Free Baselines}
\label{sec:sensor_free_baselines}

We additionally evaluate three sensor-free baselines using only the labels
available in each training fold:

\begin{itemize} [leftmargin=*]
    \item \textit{Random distribution}: We draw predictions from the
    training-fold label distribution and average the metric over 400 draws
    within each evaluation fold, with the fold index as the random seed.

    \item \textit{Rounded mean}: We assign every test instance the rounded
    mean of the training-fold labels for the task.

    \item \textit{Mode}: We assign every test instance the most frequent
    label in the training fold.
\end{itemize}

We recompute each baseline within every fold, and none uses a physiological
signal. They therefore provide label-only reference points for
interpreting model performance. A sensor-based model should be compared
against these baselines to determine whether its predictions add information
beyond the fold's label distribution. We report all three baselines, combining
the mode and rounded-mean rows when their saved predictions coincide
(\S\ref{sec:levels} discusses mode's aggregation properties).

\subsection{RQ2: Definition Codes as Model Input}
\label{sec:definition_context_model}

RQ2's second half asks whether supplying a learner's own definition of
engagement improves engagement detection. We enter that definition as an
attribute of the participant rather than as sensor input: the interview text
itself never reaches the model, only the Cognitive, Behavioral, and Emotional
codes recovered from it (Section~\ref{sec:mixed_methods_analysis}), coded as
three binary indicators per participant.

\begin{figure}[htbp]
  \centering
  \includegraphics{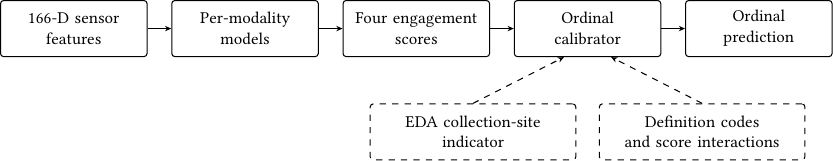}
  \caption{Definition context as model input. Within each of the six folds,
  sensor features pass through that fold's per-modality models, yielding four
  engagement scores. A held-out participant's scores are that fold's
  out-of-sample predictions for them. A second-stage ordinal calibrator receives
  those scores and the EDA collection-site indicator. The definition-conditioned
  arm additionally receives the cognitive, behavioral, and emotional codes and
  their interactions with the four scores. The codes do not enter the 166-D
  feature vector. Section~\ref{sec:definition_context_methods} describes how
  upstream scores are generated and used during calibrator validation.}
  \Description{A solid path runs from the 166-dimensional sensor feature vector
  through that fold's per-modality engagement models to four engagement scores, then
  through a second-stage ordinal calibrator to an ordinal prediction. Dashed
  inputs show that the electrodermal-activity collection-site indicator enters
  the calibrator once, and that the definition-conditioned arm additionally
  supplies cognitive, behavioral, and emotional definition codes and their
  interactions with the four scores. The definition codes do not enter the
  feature vector. Upstream fitting and selection are rebuilt inside calibrator
  validation, with the calibration arms sharing the resulting scores.}
  \label{fig:definition_context}
\end{figure}

The codes enter only a second-stage ordinal calibrator --- a five-level cumulative-link
model~\cite{mccullagh1980regression} that models the levels explicitly ---
fit over four per-modality engagement scores (fused-modality, EEG, eye, EDA)
plus the site indicator (Figure~\ref{fig:definition_context}). All three
calibration arms are fit on identical upstream inputs, the same frozen,
cross-fitted training scores and out-of-sample test scores within each outer
partition (Section~\ref{sec:definition_context_methods}), so any difference
between arms comes only from the code covariates each one adds.
These four scores are internal RQ2 covariates and are not interpreted as RQ1
sensor-ablation results.

The gating arm additionally includes an interaction between each of the three definition codes and each of the four sensor-derived scores, resulting in 12 interaction terms. A matched control --- the
four scores and the site indicator, with no codes --- supplies the reference
against which both definition arms are read. Two further baselines complete
the comparison. The fused-modality score used on its own, with no calibrator
inputs, and a constant predictor fixed at ordinal level 3.

\section{Evaluation Methodology}
\label{sec:evaluation}

We evaluate two claims under participant-independent protocols. RQ1 tests how
accurately the complete head-confined \ent configuration predicts segment-level
engagement and how performance varies across fixed classical model families.
RQ2 tests whether learners' own definitions of engagement improve
the calibration of these predictions. In both evaluations, each participant is
assigned to a single outer test fold and is excluded from all model fitting for
that fold. RQ1 trains on participants from both EDA sites and evaluates held-out
forehead participants. RQ2 uses the eighteen participants who provided explicit
engagement definitions and constructs its upstream predictions within each RQ2
training partition.

\subsection{Targets and Metrics}
\label{sec:shared_protocol}
\label{sec:metrics}

The primary target is the five-level segment rating, coded 0--4. We also conduct
a binary sensitivity analysis in which codes 3 (Engaged) and 4 (Extremely
Engaged) are positive and codes 0--2 are negative. Binary performance is
measured by macro-F1, the mean F1 across the two classes.

For ordinal prediction, let $\mathcal{C}_{\mathrm{test}}=\{c:N_c>0\}$ denote
the classes represented in a scored fold and let $N_c$ denote the number of
samples in class $c$. We report balanced 1-off accuracy and macro-MAE:
\begin{align}
\operatorname{B1Off} & =
\frac{1}{|\mathcal{C}_{\mathrm{test}}|}
\sum_{c\in\mathcal{C}_{\mathrm{test}}}\frac{1}{N_c}
\sum_{i:y_i=c}\mathbb{I}(|\hat y_i-y_i|\leq1),\\
\operatorname{MacroMAE} & =
\frac{1}{|\mathcal{C}_{\mathrm{test}}|}
\sum_{c\in\mathcal{C}_{\mathrm{test}}}\frac{1}{N_c}
\sum_{i:y_i=c}|\hat y_i-y_i|.
\end{align}
Balanced 1-off accuracy gives equal weight to each represented class and credits
predictions within one ordinal level of the target. Macro-MAE gives equal class
weight to absolute ordinal error.

\subsection{RQ1: Engagement from the Head-Confined \ent Configuration}
\label{sec:rq1_methods}

\subsubsection{Participant-Independent Evaluation}
\label{sec:cross_validation}

We use five participant-grouped outer folds. The sorted list of 30
participants is shuffled and assigned to five groups of six. Each
fold trains on the remaining 24 participants from both EDA sites and evaluates
the forehead participants in the held-out group. 

The evaluation set contains 225 segments: 24, 31, 34, 84, and 52 for ordinal
codes 0--4, respectively. The binary analysis contains 136 positive and 89
negative segments. This protocol matches the target setting of mixed-site model
development followed by forehead-only deployment. 

\subsubsection{Model Development}
\label{sec:training_procedure}

For each prediction task, we independently fit and evaluate a linear model,
LightGBM, XGBoost, AdaBoost, and Random Forest on the same 166-dimensional \ent
representation. Each family is a separately reported row, and its identity and
within-family hyperparameters remain fixed across the five outer folds. Every
family uses the same participant assignments in each fold.

All preprocessing is learned from the relevant outer training split. This
includes EDA median imputation and standardization.
The linear,
LightGBM, and Random Forest classifiers use balanced class weights. The other
classifiers use their configured defaults. Ordinal regressors produce a
continuous score that is rounded to the nearest integer and clipped to codes
0--4. We report descriptive equal-fold means and population standard deviations;
RQ1 makes no null-hypothesis significance claim.

As a secondary sensitivity analysis, we repeat the fixed-family comparison with
fold-local top-20 feature selection. Within each outer fold and model family, we
fit an initial model on the 24 training participants, rank the 166 inputs using
model-derived importance (or absolute coefficient magnitude for the linear
model), retain the 20 highest-ranked features, and refit the same family on the
reduced representation. EDA imputation, scaling, and feature ranking are learned
from the outer-fold training data only; the outer-test participants inform none
of these steps. The feature ranking is recomputed independently in every outer
fold. All five families are reported
separately, and the top-20 cutoff is fixed.

\subsection{RQ2: Learner Definitions of Engagement}
\label{sec:rq2_methods}

\subsubsection{Thematic Analysis and Definition Codes}
\label{sec:mixed_methods_analysis}

We analyzed the interview transcripts using reflexive thematic
analysis~\cite{braun2006using,braun2019reflecting}. One author conducted the
initial coding and developed an inductive coding frame. Two additional authors
reviewed the interpretations, proposed alternatives, and resolved divergent
readings through discussion. The alignment with the tripartite framework
emerged from this analysis.

Eighteen of the 24 interviews contained an explicit definition of engagement.
We represented these definitions with three overlapping participant-level
indicators: Cognitive (seven participants), Behavioral (five), and Emotional
(eight). The six interviews without an explicit definition were excluded from
the prediction analysis. 

\subsubsection{Definition-Conditioned Prediction}
\label{sec:definition_context_methods}

RQ2 uses the eighteen coded participants and their 270 segments. Six outer
folds, grouped by participant, each hold out three participants (45 segments).
All eighteen participants are evaluated once.

Within each outer fold, modality-specific models generate cross-fitted scores
for the fifteen training participants and out-of-sample scores for the three
test participants. These models use 166 mean features and AdaBoost regressors.

The second-stage ordinal calibrator receives four scores---fused modality, EEG,
eye tracking, and EDA---plus EDA site. We compare this score-calibration model
with two definition-aware variants. The first adds the three definition
indicators as main effects. The second adds twelve interactions between the four
scores and three definition indicators, for a total of 20 design columns. The
raw fused-modality score provides an additional reference.




\section{Results}
\label{sec:results}

\begin{table*}[htbp]
\centering\small
\setlength{\tabcolsep}{6pt}\renewcommand{\arraystretch}{1.12}
\caption{\textbf{Participant-independent engagement prediction with \ent.}
Every learned model receives the same 166-dimensional \ent representation.
Values are equal-fold means $\pm$ population SD across five outer folds, with
model family fixed across folds. Ordinal metrics are primary, and binary
macro-F1 is secondary. Random-distribution results average 400
draws per fold. Higher is better for balanced 1-off accuracy and binary macro-F1;
lower is better for macro-MAE. Bold marks the best point estimate in each column.}
\Description{Seven rows compare two sensor-free baselines with five classical
model families using the same fused \ent representation. AdaBoost has the highest
balanced 1-off accuracy and lowest macro-MAE. LightGBM has the highest binary
macro-F1.}
\label{tab:main_results}
\begin{tabular}{llccc}
\toprule
Group & Model & Balanced 1-off (\%) & Macro-MAE & Binary macro-F1 (\%)\\
\midrule
Sensor-free & Mode / rounded mean & 63.0 $\pm$ 6.0 & 1.320 $\pm$ 0.160 & 37.6 $\pm$ 4.4 \\
& Random distribution & 54.9 $\pm$ 3.3 & 1.520 $\pm$ 0.092 & 49.0 $\pm$ 1.1 \\
\midrule
\ent & Linear model & 57.6 $\pm$ 16.8 & 1.239 $\pm$ 0.258 & 50.5 $\pm$ 11.2 \\
& Random Forest & 67.3 $\pm$ 5.1 & 1.161 $\pm$ 0.097 & 47.2 $\pm$ 4.6 \\
& XGBoost & 70.0 $\pm$ 5.9 & 1.106 $\pm$ 0.126 & 56.7 $\pm$ 5.1 \\
& LightGBM & 72.9 $\pm$ 5.5 & 1.078 $\pm$ 0.115 & \textbf{58.9 $\pm$ 7.9} \\
& AdaBoost & \textbf{75.0 $\pm$ 7.0} & \textbf{1.043 $\pm$ 0.158} & 54.3 $\pm$ 3.0 \\
\bottomrule
\end{tabular}
\end{table*}

\subsection{RQ1: Engagement from the Head-Confined \ent Configuration}
\label{sec:rq1_results}

\subsubsection{\ent Model-Family Comparison}
\label{sec:en3_model_results}

AdaBoost has the strongest ordinal point estimates in
Table~\ref{tab:main_results}, reaching $75.0\%\pm7.0$ balanced 1-off accuracy and
$1.043\pm0.158$ macro-MAE. Relative to the mode baseline, these equal-fold means
differ by $+12.0$ percentage points in balanced 1-off accuracy and $-0.277$ in
macro-MAE. These comparisons are descriptive.

The other tree ensembles attain balanced 1-off means from $67.3\%$ to $72.9\%$,
whereas the linear model reaches $57.6\%$. LightGBM has the second-highest
ordinal estimates, with $72.9\%\pm5.5$ balanced 1-off accuracy and
$1.078\pm0.115$ macro-MAE. The ordering changes for the secondary binary target:
LightGBM has the highest macro-F1 point estimate at $58.9\%\pm7.9$, followed by
XGBoost at $56.7\%\pm5.1$ and AdaBoost at $54.3\%\pm3.0$.

\subsubsection{Sensitivity Analyses}

The binary results show that model-family rankings depend on whether engagement
is represented as a five-level ordinal outcome or collapsed into two classes.
For reference, the random-distribution binary baseline reaches
$49.0\%\pm1.1$ macro-F1.

We also refit every family after excluding the four participant-session
recordings without measured EDA from both training and evaluation. This leaves
215 test segments. AdaBoost remains the highest ordinal point estimate across
the fixed families, reaching $72.7\%\pm4.8$ balanced 1-off accuracy and
$1.066\pm0.131$ macro-MAE. This complete-case result is a descriptive
missing-data sensitivity analysis. 

The fixed-family feature-selection results in
Table~\ref{tab:rq1_feature_selection} show a family-dependent pattern. Retaining
the top 20 features improves both ordinal metrics for the linear and Random
Forest models, but worsens both for XGBoost and LightGBM. For AdaBoost, balanced
1-off accuracy increases by $0.9$ percentage points, from $75.0\%$ to $75.9\%$,
while macro-MAE increases by $0.005$, from $1.043$ to $1.048$, and binary
macro-F1 decreases by $1.3$ percentage points, from $54.3\%$ to $53.0\%$.
Feature-selection effects are therefore mixed rather than uniformly beneficial,
supporting the complete 166-dimensional representation as the primary analysis.

\begin{table*}[htbp]
\centering\small
\setlength{\tabcolsep}{6pt}
\renewcommand{\arraystretch}{1.12}
\caption{\textbf{Fixed-family performance with fold-local top-20 feature
selection.} Values are equal-fold means $\pm$ population SD across five
participant-grouped outer folds. Feature rankings are learned only from each
outer fold's training participants.}
\Description{Comparison of five model families using fold-local top-20 feature selection. Values are means and population standard deviations across five participant-grouped outer folds. AdaBoost has the highest balanced 1-off accuracy, 75.9 percent, and lowest macro-MAE, 1.048. Random Forest ranks second on both measures but has the lowest binary macro-F1, 46.7 percent. LightGBM has the highest binary macro-F1, 58.0 percent.}
\label{tab:rq1_feature_selection}
\begin{tabular}{lccc}
\toprule
Model family & Balanced 1-off (\%) & Macro-MAE & Binary macro-F1 (\%) \\
\midrule
Linear model  & 66.2 $\pm$ 10.1 & 1.141 $\pm$ 0.156 & 53.7 $\pm$ 12.7 \\
Random Forest & 72.4 $\pm$ 9.9  & 1.089 $\pm$ 0.151 & 46.7 $\pm$ 6.0 \\
XGBoost       & 65.3 $\pm$ 9.5  & 1.201 $\pm$ 0.126 & 52.6 $\pm$ 12.1 \\
LightGBM      & 67.7 $\pm$ 4.4  & 1.161 $\pm$ 0.106 & 58.0 $\pm$ 6.5 \\
AdaBoost      & 75.9 $\pm$ 3.6  & 1.048 $\pm$ 0.120 & 53.0 $\pm$ 3.0 \\
\bottomrule
\end{tabular}
\end{table*}

\begin{figure*}[htbp]
\centering
\includegraphics[width=0.58\textwidth]{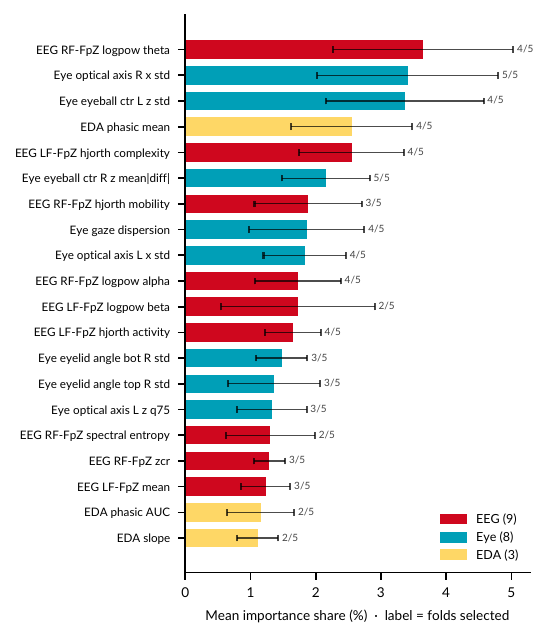}
\caption{\textbf{Feature importance underlying fixed-family AdaBoost
selection.} For the AdaBoost top-20 sensitivity arm in
Table~\ref{tab:rq1_feature_selection}, bars show the mean over five outer folds
of each feature's importance share in the initial 166-feature fit used for
ranking; error bars show $\pm1$ SD across folds. Right-hand labels give the
number of folds in which each feature entered the fold-local top 20. The 20
displayed features account for $38.6\%$ of the mean total importance mass and
comprise nine EEG, eight eye-tracking, and three EDA features. This figure
corresponds to the fixed-family AdaBoost result ($75.9\%\pm3.6$ balanced 1-off
accuracy; $1.048\pm0.120$ macro-MAE), not a nested family-selection result.}
\Description{Horizontal bar chart of 20 features ranked by their mean AdaBoost
importance share across five outer folds. EEG features are red, eye-tracking
features are teal, and EDA features are yellow. EEG RF--FpZ theta log power is
highest at about 3.65 percent and EDA slope is lowest at about 1.11 percent.
Error bars show one standard deviation, and labels report selection frequencies
from two of five through five of five folds. The modality counts are nine EEG,
eight eye-tracking, and three EDA features.}
\label{fig:top20_feature_importance}
\end{figure*}

Figure~\ref{fig:top20_feature_importance} characterizes the AdaBoost importance
ranking that drives its fold-local top-20 refit.

\subsection{RQ2: Learner Definitions of Engagement}
\label{sec:rq2_results}

\subsubsection{How Participants Define Engagement}
\label{sec:definition_themes}

Participants defined engagement through learning and understanding, sustained
attention, interest, and emotion. These themes align with the cognitive,
behavioral, and emotional dimensions of the tripartite
framework~\cite{fredricks2004school}, while showing that participants relied on
different criteria when rating their engagement.

Cognitive accounts emphasized effort and knowledge gained:
\qualquote{I think for me it would be like whether teaching me something new\ldots
In the end, I am learning something}{P18}. Behavioral accounts emphasized
sustained attention: \qualquote{As long as my thought didn't get distracted and
as long as I didn't zone out of the video\ldots that was a metric for
engagement}{P21}. Emotional accounts centered on interest and curiosity:
\qualquote{My metric is mostly focused on if the things which are being said
about it is of my interest. And when I hear about it, I want to know more about
it}{P22}.


The interviews also identified influences on engagement that were distinct from
participants' definitions: audiovisual presentation and pacing, content novelty
and jargon, prior familiarity, and structural clarity. These observations
provide context for how participants formed their segment ratings.

\subsubsection{Definition-Conditioned Calibration}
\label{sec:definition_context_results}

\begin{table}[htbp]
\centering\small
\setlength{\tabcolsep}{5pt}\renewcommand{\arraystretch}{1.12}
\caption{RQ2 ordinal prediction for eighteen coded participants under a
six-fold participant-independent protocol. Values are equal-fold means. Score
calibration is the primary comparator for the two definition-aware models;
parentheses give the number of design columns. Adding definition--score
interactions yields the strongest performance on both ordinal outcomes.}
\Description{Rows show constant code 3, raw all-modality score, score
calibration, score calibration plus definition main effects, and score
calibration plus definition-score interactions. Balanced 1-off accuracy is
60.0, 63.2, 64.6, 66.1, and 71.5 percent, respectively. Macro-MAE is 1.400,
1.216, 1.216, 1.224, and 1.124.}
\label{tab:definition_arms}
\begin{tabular}{llcc}
\toprule
Model & Inputs & Balanced 1-off (\%) & Macro-MAE\\
\midrule
Constant code 3 & No input (0) & 60.0 & 1.400\\
Raw all-modality score & Score used directly (1) & 63.2 & 1.216\\
Score calibration & Four scores + site (5) & 64.6 & 1.216\\
$+$ Definition main effects & Above + three codes (8) & 66.1 & 1.224\\
$+$ Definition--score interactions & Above + interactions (20) & 71.5 & 1.124\\
\bottomrule
\end{tabular}
\end{table}

Conditioning the modality scores on learners' definition codes produces the
strongest RQ2 performance (Table~\ref{tab:definition_arms}). The interaction
model reaches $71.5\%$ balanced 1-off accuracy and $1.124$ macro-MAE, compared
with $64.6\%$ and $1.216$ for score calibration. The fold-average improvement
is 6.9 percentage points in balanced 1-off accuracy.  The macro-MAE reduction is
$0.092$ in the fold averages. 

Adding the definition codes as main effects gives $66.1\%$ balanced 1-off
accuracy and $1.224$ macro-MAE. The larger gain from definition--score
interactions is consistent with learners' definitions modulating how the four
upstream RQ2 scores map to self-reported ratings. With eighteen coded participants, this
mechanism is an exploratory finding to test in a larger cohort.


\section{Discussion}
\label{sec:discussion}
We summarize the main findings for our two research questions
(Section~\ref{sec:discussion_findings}), discuss their implications for engagement
sensing and learning interfaces (Section~\ref{sec:discussion_implications}), and
outline the study's limitations and directions for future work
(Section~\ref{sec:discussion_limitations}).

\subsection{Main Findings}
\label{sec:discussion_findings}

\noindent\textbf{RQ1: \ent provides participant-independent feasibility evidence for head-confined engagement sensing.}
The combined EEG, eye-tracking, and forehead EDA configuration predicted
five-level engagement ratings from held-out learners after model development
using data from both EDA sites. AdaBoost achieved the strongest ordinal point
estimates: $75.0\%\pm7.0$ balanced 1-off accuracy and $1.043\pm0.158$
macro-MAE (Table~\ref{tab:main_results}). These equal-fold means exceed the mode
baseline by $12.0$ percentage points in balanced 1-off accuracy and reduce
macro-MAE by $0.277$. The gains are descriptive and support the feasibility of
the complete configuration under the tested protocol.

Model-family rankings depended on the prediction target. AdaBoost led the
five-level ordinal evaluation, whereas LightGBM achieved the highest secondary
binary macro-F1 at $58.9\%\pm7.9$. AdaBoost also retained the strongest ordinal
point estimates in the complete-case analysis, reaching $72.7\%\pm4.8$
balanced 1-off accuracy and $1.066\pm0.131$ macro-MAE after recordings without
measured EDA were excluded. Feature selection had mixed effects across model
families (Table~\ref{tab:rq1_feature_selection}), supporting the complete
166-dimensional representation as the primary analysis.

\noindent\textbf{RQ2: Learners' definitions provide promising context for mapping sensor-derived scores to engagement ratings.}
Learners described engagement through understanding, sustained attention,
interest, and emotion. These accounts span the cognitive, behavioral, and
emotional dimensions of engagement and indicate that learners can draw on
different criteria when assigning the same overall rating. In the exploratory
analysis of eighteen participants, adding definition codes and their
interactions with four frozen sensor-derived scores increased balanced 1-off
accuracy from $64.6\%$ to $71.5\%$ and reduced macro-MAE from $1.216$ to
$1.124$ relative to matched score calibration
(Table~\ref{tab:definition_arms}). Adding definition main effects alone yielded
$66.1\%$ balanced 1-off accuracy and $1.224$ macro-MAE. The stronger results
with interactions are consistent with the hypothesis that definitions help
condition how sensor-derived scores map to reported engagement. RQ2 uses its
own cohort and evaluation protocol, its scores should be interpreted within
that comparison rather than directly against RQ1.

\subsection{Implications}
\label{sec:discussion_implications}

\subsubsection{Designing Engagement Sensing Around Head-Worn Devices}
\ent provides a concrete sensing layout for combining EEG, eye tracking, and EDA
within a head-worn configuration. Prior engagement systems such as SenseSeek
and Effecti-Net combine EEG or eye tracking with EDA recorded at the wrist or
fingers~\cite{ji2025senseseek,dwivedi2024effecti}, while Galea and HMDBioPad
demonstrate co-located head-worn physiological
instrumentation~\cite{bernal2022galea,wan2022hmdbiopad}. \ent complements these
systems by evaluating the complete head-confined combination against
segment-level ratings from held-out learners.

This result motivates further integration of the three modalities into a
shared form factor. Such integration could reduce the number of separately
worn components and simplify fitting and synchronization. These are design
opportunities to evaluate: the present configuration uses three devices, and
its predictive performance does not establish comfort or ease of everyday use.
The findings provide a basis for investigating whether a more integrated
device can preserve useful engagement estimates while reducing wearing and
setup demands.

\subsubsection{Personalizing the Interpretation of Engagement}
RQ2 connects engagement sensing with the multidimensional account of engagement
in educational psychology~\cite{fredricks2004school,benEliyahu2018multidimensionality}.
A single self-report can reflect several dimensions, and learners may emphasize
them differently. This matters because physiological measurements acquire
interpretive meaning through the labels with which they are
associated~\cite{gao2020n,gao2022understanding}. Learners' descriptions make some
of those rating criteria explicit and offer context for interpreting model
outputs.

The resulting design implication is to include learners in defining what an
engagement estimate represents. A future interface could briefly elicit whether
a learner associates engagement with understanding, sustained attention,
interest, or a combination of these, and use that information when mapping
sensor-derived scores to ratings. The learner could review and update these
criteria as tasks or goals change. Our interaction results motivate this
approach, although prospective elicitation and its stability remain to be
tested. The present codes were derived from interviews, they were not inferred
as psychological traits from sensor data.

Definition context could also inform the design of learning support. For
example, an interface could invite a learner who emphasizes understanding to
review a difficult concept, or offer a pacing adjustment to someone concerned
with sustaining attention. These are candidate interactions for future studies.
The current models predict reported engagement and do not establish the cause
of a low rating or which intervention would help.

\subsubsection{Evaluating Performance Across Engagement Levels}
\label{sec:levels}
Our findings emphasize the importance of evaluating less frequent engagement
states. Ratings were concentrated in the engaged categories, making aggregate
performance sensitive to the label distribution. The mode baseline illustrates
this issue: always predicting code 3 receives within-one credit for codes
2--4 and attains $63.0\%$ balanced 1-off accuracy in RQ1. A within-one score
therefore requires interpretation alongside error magnitude and an appropriate
sensor-free reference.

Macro-MAE averages absolute error equally across represented classes,
preventing frequent ratings from dominating the
measure~\cite{baccianella2009evaluation}. Balanced 1-off accuracy applies the
same class weighting to agreement within one rating level. We encourage
engagement-sensing studies to report both measures alongside label
distributions, per-class results, and constant baselines. For learning
interfaces, model selection should also reflect the intended response: tracking
graded changes and triggering a binary alert require different evaluations,
as the change in model-family rankings between our ordinal and binary tasks
illustrates.

\subsection{Limitations and Future Work}
\label{sec:discussion_limitations}

\subsubsection{Sample Size and Generalization}
The study recruited thirty participants from one university community. RQ1
evaluated fifteen forehead participants, while RQ2 included eighteen participants
who supplied explicit engagement definitions. Repeated segments increase the
number of observations but do not increase the number of independent learners.
The reported differences are descriptive, and these sample sizes limit claims
about performance in broader populations. Participant-disjoint evaluation
addresses transfer to unseen learners within this study, while the shared set
of four educational videos leaves transfer to unseen materials untested.
Larger studies should include more diverse learners and learning activities,
with evaluation that holds out both participants and content.

\subsubsection{Sensor Placement, Contribution, and Missing Data}
EDA placement was confounded with collection batch: the first fifteen
participants wore wrist EDA and the next fifteen wore forehead EDA. No
participant wore both. Consequently, the study cannot isolate placement effects
or quantify a performance trade-off between head-confined and distributed
sensing. Moreover, every learned RQ1 model received the complete \ent
representation, so its results do not establish the contribution of each
modality. A within-participant study recording both EDA sites, with
counterbalanced conditions and prespecified sensor ablations, would address
these questions.

Four participant-session recordings lacked usable EDA and required fold-local
imputation in the primary analysis. The complete-case analysis retained
AdaBoost's ordinal ranking, but excluding those recordings changed both the
training and test composition. It therefore cannot isolate the effect of
imputation or establish robustness to sensor failure. Future evaluation should
test predefined sensor-loss conditions and signal-quality variation.

\subsubsection{Retrospective Ratings and Temporal Resolution}
Participants rated segments retrospectively after each video. This preserved
uninterrupted viewing, but ratings may reflect recall and impressions formed
over the whole video. The model predicts these reports, which are not a direct
measure of learning achievement. In addition, features summarize the first
120 seconds of segments lasting approximately 140--155 seconds. This partial
coverage and temporal aggregation may obscure changes within a segment.
Future work should compare labeling schedules and analysis-window lengths,
assess agreement with complementary measures of engagement and learning, and
test the temporal resolution needed for timely support.

\subsubsection{Definition Elicitation and Model Complexity}
RQ2 represents eighteen participant profiles using three overlapping definition
indicators. The full calibration design contains twenty columns, including
three definition main effects and twelve definition--score interactions. This
complexity relative to the number of learners limits confidence in the
stability of the observed gain. Holding upstream scores fixed makes the
comparison specific to the added definition terms, but does not establish that
the gain will reproduce or that definitions explain it independently of other
participant or cohort differences.

Definitions were also elicited after the task and may reflect the completed
study experience. Reducing interview accounts to three indicators loses
variation within each category, and their stability across learning contexts
was not assessed. A larger, preregistered study should elicit definitions before
viewing, evaluate a smaller prespecified interaction set, and repeat elicitation
across tasks. Any model selection should occur within participant-disjoint
training folds, followed by evaluation on new learners and materials.

\subsubsection{From Prediction to Everyday Learning Support}
The evaluation concerns offline prediction during seated video viewing. It does
not establish long-term wearability, performance during everyday movement, or
the benefit of responding to predicted engagement. Future work should evaluate
comfort, setup burden, signal reliability, and prediction latency in everyday
learning settings. An interactive study could then compare sensing-informed
support with a suitable control, measuring learning outcomes and interruption
burden as well as prediction quality. This would test whether the sensing and
definition context investigated here translate into useful support for learners.
\section{Conclusion and Future Work}
\label{sec:conclusion}

\ent co-locates EEG, eye tracking, and forehead EDA as a head-confined sensing
configuration. 
AdaBoost using the complete \ent representation achieves
$75.0\%\pm7.0$ balanced 1-off accuracy and $1.043\pm0.158$ macro-MAE on
held-out forehead participants. Its balanced 1-off fold mean is $12.0$
percentage points above mode. These descriptive results provide preliminary
feasibility evidence for the complete head-confined configuration under the
tested protocol. Because every learned RQ1 model uses the full \ent
representation, this comparison does not estimate the contribution of any
individual sensor. Controlled ablations in a larger cohort remain future work.

Learners' exit-interview accounts are an initial, exploratory check on what
that sensing result predicts. In a coded-subsample analysis, the
interaction arm over frozen, cross-fitted upstream scores attains $71.5\%$
balanced 1-off accuracy versus $64.6\%$ for score calibration, with
macro-MAE decreasing
from $1.216$ to $1.124$. This result is consistent with the construct \ent's
models predict being something learners can name, and naming it sharpening
the prediction (see \S\ref{sec:discussion}).

Future work should test sensor attribution, transfer to new learning contexts,
and definition-conditioned calibration in larger cohorts
(\S\ref{sec:discussion}). Taken together, these results indicate that a
fully head-confined sensor set can predict self-reported engagement, with
learners' own accounts of what engagement means to them providing a
source of corroborating context sensing design has not yet drawn on.

\subsection*{Data and Code Availability}
Analysis code and de-identified derived features supporting the reported
results will be released upon request.

\bibliographystyle{ACM-Reference-Format}
\bibliography{references}

@String{Computing = "Computing" }

@String{Computer = "{IEEE} Computer" }

@String{Academic = "Academic Press" }

@String{Springer = "Springer-Verlag" }

@BOOK{test,
   author = "Donald E. Knuth",
   title = "Seminumerical Algorithms",
   volume = 2,
   series = "The Art of Computer Programming",
   publisher = "Addison-Wesley",
   address = "Reading, MA",
   edition = "2nd",
   month = "10~" # jan,
   year = "1981",
}

@string{IEEE_J_CYB  = "IEEE Transactions on Cybernetics"}

@string{IEEE_J_PAMI = "IEEE Transactions on Pattern Analysis and Machine Intelligence"}

@string{IEEE_J_PROC = "Proceedings of the IEEE"}

@string{IEEE_J_AFFC = "IEEE Transactions on Affective Computing"}

@article{zheng2019emotionmeter,
  author  = {W.-L. Zheng and W. Liu and Y. Lu and B.-L. Lu and A. Cichocki},
  journal = IEEE_J_CYB,
  title   = {{EmotionMeter}: A Multimodal Framework for Recognizing Human Emotions},
  year    = {2019},
  volume  = {49},
  number  = {3},
  pages   = {1110--1122},
  doi     = {10.1109/TCYB.2018.2797176},
  url     = {https://doi.org/10.1109/TCYB.2018.2797176},
  issn    = {2168-2267}
}

@article{doherty2018engagement,
  title={Engagement in HCI: conception, theory and measurement},
  author={Doherty, Kevin and Doherty, Gavin},
  journal={ACM computing surveys (CSUR)},
  volume={51},
  number={5},
  pages={1--39},
  year={2018},
  publisher={ACM New York, NY, USA},
  doi     = {10.1145/3234149},
  url     = {https://doi.org/10.1145/3234149}
}

@article{mccullagh1980regression,
  author  = {McCullagh, Peter},
  title   = {Regression Models for Ordinal Data},
  journal = {Journal of the Royal Statistical Society: Series B (Methodological)},
  year    = {1980},
  volume  = {42},
  number  = {2},
  pages   = {109--127},
  doi     = {10.1111/j.2517-6161.1980.tb01109.x}
}

@article{adaboost1997,
  author  = {Freund, Yoav and Schapire, Robert E.},
  title   = {A Decision-Theoretic Generalization of On-Line Learning and an
             Application to Boosting},
  journal = {J. Comput. Syst. Sci.},
  year    = {1997},
  volume  = {55},
  number  = {1},
  pages   = {119--139},
  doi     = {10.1006/jcss.1997.1504},
  url     = {https://doi.org/10.1006/jcss.1997.1504}
}

@article{attentivu,
  title    = {{{AttentivU}}: {{An EEG-based}} Closed-Loop Biofeedback System for Real-Time Monitoring and Improvement of Engagement for Personalized Learning},
  author   = {Kosmyna, Nataliya and Maes, Pattie},
  year     = {2019},
  journal  = {Sensors},
  volume   = {19},
  number   = {23},
  pages    = {5200},
  issn     = {1424-8220},
  doi      = {10.3390/s19235200},
  url      = {https://www.mdpi.com/1424-8220/19/23/5200},
  pubmedid = {31783646}
}

@article{baltrusaitis2019multimodal,
  author  = {Baltru{\v{s}}aitis, Tadas and Ahuja, Chaitanya and Morency, Louis-Philippe},
  title   = {Multimodal Machine Learning: A Survey and Taxonomy},
  journal = IEEE_J_PAMI,
  volume  = {41},
  number  = {2},
  pages   = {423--443},
  year    = {2019},
  doi     = {10.1109/TPAMI.2018.2798607},
  url     = {https://doi.org/10.1109/TPAMI.2018.2798607}
}

@inproceedings{baccianella2009evaluation,
  title     = {Evaluation Measures for Ordinal Regression},
  author    = {Baccianella, Stefano and Esuli, Andrea and Sebastiani, Fabrizio},
  booktitle = {Proc. 9th Int. Conf. Intell. Syst. Des. Appl. (ISDA)},
  pages     = {283--287},
  year      = {2009},
  publisher = {IEEE},
  address   = {Piscataway, NJ, USA},
  doi       = {10.1109/ISDA.2009.230},
  url       = {https://doi.org/10.1109/ISDA.2009.230}
}

@inproceedings{babaei2021critique,
  author    = {Babaei, Ebrahim and Tag, Benjamin and Dingler, Tilman and Velloso, Eduardo},
  title     = {A Critique of Electrodermal Activity Practices at {CHI}},
  booktitle = {Proceedings of the 2021 {CHI} Conference on Human Factors in Computing Systems},
  year      = {2021},
  pages     = {1--14},
  publisher = {Association for Computing Machinery},
  address   = {New York, NY, USA},
  doi       = {10.1145/3411764.3445370},
  url       = {https://doi.org/10.1145/3411764.3445370}
}

@article{beatty1982task,
  author  = {Beatty, Jackson},
  title   = {Task-Evoked Pupillary Responses, Processing Load, and the Structure of Processing Resources},
  journal = {Psychol. Bull.},
  volume  = {91},
  number  = {2},
  pages   = {276--292},
  year    = {1982},
  doi     = {10.1037/0033-2909.91.2.276},
  url     = {https://doi.org/10.1037/0033-2909.91.2.276}
}

@article{benEliyahu2018multidimensionality,
  author  = {Ben-Eliyahu, Adar and Moore, Debra and Dorph, Rena and Schunn, Christian D.},
  title   = {Investigating the multidimensionality of engagement: Affective, behavioral, and cognitive engagement across science activities and contexts},
  journal = {Contemp. Educ. Psychol.},
  volume  = {53},
  pages   = {87--105},
  year    = {2018},
  doi     = {10.1016/j.cedpsych.2018.01.002},
  url     = {https://doi.org/10.1016/j.cedpsych.2018.01.002}
}

@inproceedings{bernal2022galea,
  title        = {Galea: A physiological sensing system for behavioral research in Virtual Environments},
  author       = {Bernal, Guillermo and Hidalgo, Nelson and Russomanno, Conor and Maes, Pattie},
  booktitle    = {2022 {IEEE} Conf. Virtual Reality 3D User Interfaces ({VR})},
  pages        = {66--76},
  year         = {2022},
  publisher    = {IEEE},
  address      = {Piscataway, NJ, USA},
  organization = {IEEE},
  doi     = {10.1109/vr51125.2022.00024},
  url     = {https://doi.org/10.1109/vr51125.2022.00024}
}

@article{booth2023engagement,
  title     = {Engagement detection and its applications in learning: a tutorial and selective review},
  author    = {Booth, Brandon M and Bosch, Nigel and D’Mello, Sidney K},
  journal   = IEEE_J_PROC,
  volume    = {111},
  number    = {10},
  pages     = {1398--1422},
  year      = {2023},
  publisher = {IEEE},
  doi     = {10.1109/jproc.2023.3309560},
  url     = {https://doi.org/10.1109/jproc.2023.3309560}
}

@book{boucsein2012electrodermal,
  author    = {Boucsein, Wolfram},
  title     = {Electrodermal Activity},
  publisher = {Springer},
  address   = {New York},
  edition   = {2nd},
  year      = {2012},
  doi     = {10.1007/978-1-4614-1126-0},
  url     = {https://doi.org/10.1007/978-1-4614-1126-0}
}

@article{bradley1994measuring,
  title     = {Measuring emotion: The self-assessment manikin and the semantic differential},
  author    = {Bradley, Margaret M and Lang, Peter J},
  journal   = {J. Behav. Ther. Exp. Psychiatry},
  volume    = {25},
  number    = {1},
  pages     = {49--59},
  year      = {1994},
  publisher = {Elsevier},
  doi     = {10.1016/0005-7916(94)90063-9},
  url     = {https://doi.org/10.1016/0005-7916(94)90063-9}
}

@article{braun2006using,
  author  = {Braun, Virginia and Clarke, Victoria},
  title   = {Using Thematic Analysis in Psychology},
  journal = {Qual. Res. Psychol.},
  volume  = {3},
  number  = {2},
  pages   = {77--101},
  year    = {2006},
  doi     = {10.1191/1478088706qp063oa},
  url     = {https://doi.org/10.1191/1478088706qp063oa}
}

@article{braun2019reflecting,
  author  = {Braun, Virginia and Clarke, Victoria},
  title   = {Reflecting on Reflexive Thematic Analysis},
  journal = {Qual. Res. Sport Exerc. Health},
  volume  = {11},
  number  = {4},
  pages   = {589--597},
  year    = {2019},
  doi     = {10.1080/2159676X.2019.1628806},
  url     = {https://doi.org/10.1080/2159676X.2019.1628806}
}

@article{brishtel2020mind,
  author  = {Brishtel, Iuliia and Khan, Anam Ahmad and Schmidt, Thomas and Dingler, Tilman and Ishimaru, Shoya and Dengel, Andreas},
  title   = {Mind Wandering in a Multimodal Reading Setting: Behavior Analysis \& Automatic Detection Using Eye-Tracking and an {EDA} Sensor},
  journal = {Sensors},
  volume  = {20},
  number  = {9},
  pages   = {2546},
  year    = {2020},
  doi     = {10.3390/s20092546},
  url     = {https://doi.org/10.3390/s20092546}
}

@article{bustos2022wearables,
  title     = {Wearables for engagement detection in learning environments: A review},
  author    = {Bustos-Lopez, Maritza and Cruz-Ramirez, Nicandro and Guerra-Hernandez, Alejandro and S{\'a}nchez-Morales, Laura Nely and Cruz-Ramos, Nancy Aracely and Alor-Hernandez, Giner},
  journal   = {Biosensors},
  volume    = {12},
  number    = {7},
  pages     = {509},
  year      = {2022},
  publisher = {MDPI},
  doi     = {10.3390/bios12070509},
  url     = {https://doi.org/10.3390/bios12070509}
}

@article{casson2019wearable,
  title     = {Wearable {EEG} and Beyond},
  author    = {Casson, Alexander J},
  journal   = {Biomed. Eng. Lett.},
  volume    = {9},
  number    = {1},
  pages     = {53--71},
  year      = {2019},
  publisher = {Springer},
  doi     = {10.1007/s13534-018-00093-6},
  url     = {https://doi.org/10.1007/s13534-018-00093-6}
}

@inproceedings{de2019engaged,
  title     = {“Engaged Faces”: Measuring and Monitoring Student Engagement from Face and Gaze Behavior},
  author    = {De Carolis, Berardina and D'Errico, Francesca and Macchiarulo, Nicola and Palestra, Giuseppe},
  booktitle = {IEEE/WIC/ACM Int. Conf. Web Intell. Companion Volume},
  pages     = {80--85},
  year      = {2019},
  publisher = {Association for Computing Machinery},
  address   = {New York, NY, USA},
  doi     = {10.1145/3358695.3361748},
  url     = {https://doi.org/10.1145/3358695.3361748}
}

@inproceedings{dhall2018emotiw,
  author    = {Dhall, Abhinav and Kaur, Amanjot and Goecke, Roland and Gedeon, Tom},
  title     = {{EmotiW} 2018: Audio-Video, Student Engagement and Group-Level Affect Prediction},
  booktitle = {Proc. 20th ACM Int. Conf. Multimodal Interact. (ICMI)},
  pages     = {653--656},
  year      = {2018},
  publisher = {ACM},
  address   = {New York, NY, USA},
  doi       = {10.1145/3242969.3264993},
  url       = {https://doi.org/10.1145/3242969.3264993}
}

@article{diLascio2018engagement,
  author     = {Di Lascio, Elena and Gashi, Shkurta and Santini, Silvia},
  title      = {Unobtrusive Assessment of Students' Emotional Engagement during Lectures Using Electrodermal Activity Sensors},
  year       = {2018},
  issue_date = {September 2018},
  publisher  = {Association for Computing Machinery},
  address    = {New York, NY, USA},
  volume     = {2},
  number     = {3},
  url        = {https://doi.org/10.1145/3264913},
  doi        = {10.1145/3264913},
  journal    = {Proc. ACM Interact. Mob. Wearable Ubiquitous Technol.},
  month      = sep,
  articleno  = {103},
  numpages   = {21}
}

@inproceedings{dwivedi2024effecti,
  title     = {{Effecti-Net}: A multimodal framework and database for educational content effectiveness analysis},
  author    = {Dwivedi, Deep and Garg, Ritik and Baghel, Shiva and Thareja, Rushil and Kulshrestha, Ritvik and Mohania, Mukesh and Shukla, Jainendra},
  booktitle = {Proc. 14th Learning Analytics Knowl. Conf. ({LAK})},
  pages     = {667--677},
  year      = {2024},
  publisher = {Association for Computing Machinery},
  address   = {New York, NY, USA},
  doi       = {10.1145/3636555.3636928},
  url       = {https://doi.org/10.1145/3636555.3636928}
}

@article{fredricks2004school,
  title     = {School engagement: Potential of the concept, state of the evidence},
  author    = {Fredricks, Jennifer A and Blumenfeld, Phyllis C and Paris, Alison H},
  journal   = {Rev. Educ. Res.},
  volume    = {74},
  number    = {1},
  pages     = {59--109},
  year      = {2004},
  publisher = {Sage Publications},
  doi     = {10.3102/00346543074001059},
  url     = {https://doi.org/10.3102/00346543074001059}
}

@incollection{fredricks2012measurement,
  title     = {The measurement of student engagement: A comparative analysis of various methods and student self-report instruments},
  author    = {Fredricks, Jennifer A and McColskey, Wendy},
  booktitle = {Handbook of research on student engagement},
  pages     = {763--782},
  year      = {2012},
  publisher = {Springer},
  address   = {New York, NY, USA},
  doi     = {10.1007/978-1-4614-2018-7_37},
  url     = {https://doi.org/10.1007/978-1-4614-2018-7_37}
}

@misc{frenzbrainband,
  year         = {[n.\,d.]},
  author       = {{Earable Inc.}},
  title        = {{FRENZ Brainband}},
  howpublished = {\url{https://frenzband.com/products/frenz-brainband}},
  note         = {Accessed: 2026-01-29}
}

@article{gao2020n,
  title     = {n-gage: Predicting in-class emotional, behavioural and cognitive engagement in the wild},
  author    = {Gao, Nan and Shao, Wei and Rahaman, Mohammad Saiedur and Salim, Flora D},
  journal   = {Proc. ACM Interact. Mob. Wearable Ubiquitous Technol.},
  volume    = {4},
  number    = {3},
  pages     = {1--26},
  year      = {2020},
  publisher = {ACM New York, NY, USA},
  doi       = {10.1145/3411813}
}

@article{gao2022understanding,
  title     = {Understanding occupants’ behaviour, engagement, emotion, and comfort indoors with heterogeneous sensors and wearables},
  author    = {Gao, Nan and Marschall, Max and Burry, Jane and Watkins, Simon and Salim, Flora D},
  journal   = {Sci. Data},
  volume    = {9},
  number    = {1},
  pages     = {261},
  year      = {2022},
  publisher = {Nature Publishing Group UK London},
  doi     = {10.1038/s41597-022-01347-w},
  url     = {https://doi.org/10.1038/s41597-022-01347-w}
}

@misc{gao2023critiquing,
  title         = {Critiquing Self-report Practices for Human Mental and Wellbeing Computing at {Ubicomp}},
  author        = {Gao, Nan and Ananthan, Soundariya and Yu, Chun and Wang, Yuntao and Salim, Flora D},
  year          = {2023},
  eprint        = {2311.15496},
  archivePrefix = {arXiv},
  primaryClass  = {cs.HC},
  doi           = {10.48550/arXiv.2311.15496}
}

@article{giannakos2019multimodal,
  author  = {Giannakos, Michail N. and Sharma, Kshitij and Pappas, Ilias O. and Kostakos, Vassilis and Velloso, Eduardo},
  title   = {Multimodal Data as a Means to Understand the Learning Experience},
  journal = {Int. J. Inf. Manage.},
  volume  = {48},
  pages   = {108--119},
  year    = {2019},
  doi     = {10.1016/j.ijinfomgt.2019.02.003},
  url     = {https://doi.org/10.1016/j.ijinfomgt.2019.02.003}
}

@inproceedings{guo2014video,
  title     = {How Video Production Affects Student Engagement: An Empirical Study of {MOOC} Videos},
  author    = {Guo, Philip J and Kim, Juho and Rubin, Rob},
  booktitle = {Proc. 1st ACM Conf. Learn. @ Scale},
  pages     = {41--50},
  year      = {2014},
  publisher = {Association for Computing Machinery},
  address   = {New York, NY, USA},
  doi       = {10.1145/2556325.2566239}
}

@misc{gupta2016daisee,
  author        = {Gupta, Abhay and D'Cunha, Arjun and Awasthi, Kamal and Balasubramanian, Vineeth},
  title         = {{DAiSEE}: Towards User Engagement Recognition in the Wild},
  year          = {2016},
  eprint        = {1609.01885},
  archivePrefix = {arXiv},
  primaryClass  = {cs.CV},
  doi           = {10.48550/arXiv.1609.01885}
}

@article{Hossain2022Comparison,
  title   = {Comparison of Electrodermal Activity from Multiple Body Locations Based on Standard {EDA} Indices' Quality and Robustness against Motion Artifact},
  author  = {Hossain, Md-Billal and Kong, Youngsun and Posada-Quintero, Hugo F. and Chon, Ki H.},
  journal = {Sensors},
  year    = {2022},
  volume  = {22},
  number  = {9},
  pages   = {3177},
  doi     = {10.3390/s22093177},
  url     = {https://doi.org/10.3390/s22093177}
}

@article{ji2025senseseek,
  title     = {{SenseSeek} Dataset: Multimodal Sensing to Study Information Seeking Behaviors},
  author    = {Ji, Kaixin and Hettiachchi, Danula and Scholer, Falk and Salim, Flora D and Spina, Damiano},
  journal   = {Proc. ACM Interact. Mob. Wearable Ubiquitous Technol.},
  volume    = {9},
  number    = {3},
  pages     = {1--29},
  year      = {2025},
  publisher = {ACM New York, NY, USA},
  doi       = {10.1145/3749501}
}

@book{kahneman1973attention,
  author    = {Kahneman, Daniel},
  title     = {Attention and Effort},
  publisher = {Prentice-Hall},
  address   = {Englewood Cliffs, NJ},
  series    = {Prentice-Hall Series in Experimental Psychology},
  year      = {1973}
}

@article{kleeva2024dryeeg,
  author  = {Kleeva, Daria and Ninenko, Ivan and Lebedev, Mikhail A.},
  title   = {Resting-State {EEG} Recorded with Gel-Based vs. Consumer Dry Electrodes: Spectral Characteristics and Across-Device Correlations},
  journal = {Front. Neurosci.},
  volume  = {18},
  pages   = {1326139},
  year    = {2024},
  doi     = {10.3389/fnins.2024.1326139},
  url     = {https://doi.org/10.3389/fnins.2024.1326139}
}

@article{kleckner2018simple,
  author  = {Kleckner, Ian R. and Jones, Rebecca M. and Wilder-Smith, Oliver and Wormwood, Jolie B. and Akcakaya, Murat and Quigley, Karen S. and Lord, Catherine and Goodwin, Matthew S.},
  title   = {Simple, Transparent, and Flexible Automated Quality Assessment Procedures for Ambulatory Electrodermal Activity Data},
  journal = {IEEE Transactions on Biomedical Engineering},
  year    = {2018},
  volume  = {65},
  number  = {7},
  pages   = {1460--1467},
  doi     = {10.1109/TBME.2017.2758643},
  url     = {https://doi.org/10.1109/TBME.2017.2758643}
}

@article{klimesch1999eeg,
  author  = {Klimesch, Wolfgang},
  title   = {{EEG} Alpha and Theta Oscillations Reflect Cognitive and Memory Performance: A Review and Analysis},
  journal = {Brain Res. Rev.},
  volume  = {29},
  number  = {2-3},
  pages   = {169--195},
  year    = {1999},
  doi     = {10.1016/S0165-0173(98)00056-3},
  url     = {https://doi.org/10.1016/S0165-0173(98)00056-3}
}

@article{knierim2025advancing,
  title     = {Advancing Wearable {BCI}: Headphone {EEG} for Cognitive Load Detection in Lab and Field},
  author    = {Knierim, Michael T and Zimny, Christian and Ivucic, Gabriel and R{\"o}ddiger, Tobias},
  journal   = {Proc. ACM Interact. Mob. Wearable Ubiquitous Technol.},
  volume    = {9},
  number    = {1},
  pages     = {1--26},
  year      = {2025},
  publisher = {ACM New York, NY, USA},
  doi       = {10.1145/3712283}
}

@article{koelstra2011deap,
  title     = {{DEAP}: A Database for Emotion Analysis Using Physiological Signals},
  author    = {Koelstra, Sander and M{\"u}hl, Christian and Soleymani, Mohammad and Lee, Jong-Seok and Yazdani, Ashkan and Ebrahimi, Touradj and Pun, Thierry and Nijholt, Anton and Patras, Ioannis},
  journal   = IEEE_J_AFFC,
  volume    = {3},
  number    = {1},
  pages     = {18--31},
  year      = {2012},
  doi       = {10.1109/T-AFFC.2011.15},
  url       = {https://doi.org/10.1109/T-AFFC.2011.15},
  publisher = {IEEE}
}

@article{krosnick1999survey,
  title     = {Survey research},
  author    = {Krosnick, Jon A},
  journal   = {Annu. Rev. Psychol.},
  volume    = {50},
  number    = {1},
  pages     = {537--567},
  year      = {1999},
  publisher = {Annual Reviews},
  doi     = {10.1146/annurev.psych.50.1.537},
  url     = {https://doi.org/10.1146/annurev.psych.50.1.537}
}

@article{lei2018relationship,
  author  = {Lei, Hao and Cui, Yunhuo and Zhou, Wenye},
  year    = {2018},
  month   = mar,
  pages   = {517--528},
  title   = {Relationships between student engagement and academic achievement: A meta-analysis},
  volume  = {46},
  number  = {3},
  journal = {Soc. Behav. Pers.},
  doi     = {10.2224/sbp.7054},
  url     = {https://doi.org/10.2224/sbp.7054}
}

@inproceedings{lightgbm2017,
  author    = {Ke, Guolin and Meng, Qi and Finley, Thomas and Wang, Taifeng
               and Chen, Wei and Ma, Weidong and Ye, Qiwei and Liu, Tie{-}Yan},
  title     = {{LightGBM}: {A} Highly Efficient Gradient Boosting Decision Tree},
  booktitle = {Adv. Neural Inf. Process. Syst. 30 (NIPS)},
  year      = {2017},
  pages     = {3146--3154},
  publisher = {Curran Associates, Inc.},
  address   = {Red Hook, NY, USA},
  url       = {https://proceedings.neurips.cc/paper/2017/hash/6449f44a102fde848669bdd9eb6b76fa-Abstract.html}
}

@inproceedings{long2024multimodal,
  author    = {Long, Xingyu and Mayer, Sven and Chiossi, Francesco},
  title     = {Multimodal Detection of External and Internal Attention in Virtual Reality Using {EEG} and Eye Tracking Features},
  year      = {2024},
  isbn      = {9798400709982},
  publisher = {Association for Computing Machinery},
  address   = {New York, NY, USA},
  url       = {https://doi.org/10.1145/3670653.3670657},
  doi       = {10.1145/3670653.3670657},
  booktitle = {Proc. Mensch und Computer 2024},
  pages     = {29--43},
  numpages  = {15},
  location  = {Karlsruhe, Germany},
  series    = {MuC '24}
}

@inproceedings{matton2023contrastive,
  title     = {Contrastive learning of electrodermal activity representations for stress detection},
  author    = {Matton, Katie and Lewis, Robert and Guttag, John and Picard, Rosalind},
  booktitle = {Conf. Health, Inference, Learn.},
  pages     = {410--426},
  year      = {2023},
  volume    = {209},
  series    = {Proceedings of Machine Learning Research},
  publisher = {PMLR},
  address   = {Cambridge, MA, USA},
  url       = {https://proceedings.mlr.press/v209/matton23a.html}
}

@article{monkaresi2016automated,
  title     = {Automated detection of engagement using video-based estimation of facial expressions and heart rate},
  author    = {Monkaresi, Hamed and Bosch, Nigel and Calvo, Rafael A and D'Mello, Sidney K},
  journal   = IEEE_J_AFFC,
  volume    = {8},
  number    = {1},
  pages     = {15--28},
  year      = {2017},
  publisher = {IEEE},
  doi     = {10.1109/taffc.2016.2515084},
  url     = {https://doi.org/10.1109/taffc.2016.2515084}
}

@article{montgomery2024validating,
  title     = {Validating {EmotiBit}, an open-source multi-modal sensor for capturing research-grade physiological signals from anywhere on the body},
  author    = {Montgomery, Sean M. and Nair, Nitin and Chen, Phoebe and Dikker, Suzanne},
  journal   = {Measurement: Sensors},
  volume    = {32},
  pages     = {101075},
  year      = {2024},
  doi       = {10.1016/j.measen.2024.101075},
  url       = {https://doi.org/10.1016/j.measen.2024.101075},
  publisher = {Elsevier}
}

@article{pope1995biocybernetic,
  author  = {Pope, Alan T. and Bogart, Edward H. and Bartolome, Debbie S.},
  title   = {Biocybernetic system evaluates indices of operator engagement in automated task},
  journal = {Biol. Psychol.},
  volume  = {40},
  number  = {1-2},
  pages   = {187--195},
  year    = {1995},
  month   = may,
  doi     = {10.1016/0301-0511(95)05116-3},
  url     = {https://doi.org/10.1016/0301-0511(95)05116-3},
  pmid    = {7647180}
}

@misc{pupillabsneon,
  year         = {[n.\,d.]},
  author       = {{Pupil Labs}},
  title        = {{Neon Eye Tracking System}},
  howpublished = {\url{https://pupil-labs.com/products/neon/specs}},
  note         = {Accessed: 2026-01-29}
}

@article{ramirez2021eeg,
  title     = {{EEG}-based tool for prediction of university students’ cognitive performance in the classroom},
  author    = {Ram{\'\i}rez-Moreno, Mauricio A and D{\'\i}az-Padilla, Mariana and Valenzuela-G{\'o}mez, Karla D and Vargas-Mart{\'\i}nez, Adriana and Tud{\'o}n-Mart{\'\i}nez, Juan C and Morales-Menendez, Rub{\'e}n and Ram{\'\i}rez-Mendoza, Ricardo A and P{\'e}rez-Henr{\'\i}quez, Blas L and Lozoya-Santos, Jorge de J},
  journal   = {Brain Sci.},
  volume    = {11},
  number    = {6},
  pages     = {698},
  year      = {2021},
  publisher = {MDPI},
  doi     = {10.3390/brainsci11060698},
  url     = {https://doi.org/10.3390/brainsci11060698}
}

@article{randomforest2001,
  author  = {Breiman, Leo},
  title   = {Random Forests},
  journal = {Mach. Learn.},
  year    = {2001},
  volume  = {45},
  number  = {1},
  pages   = {5--32},
  doi     = {10.1023/A:1010933404324},
  url     = {https://doi.org/10.1023/A:1010933404324}
}

@article{rayner1998eye,
  title   = {Eye movements in reading and information processing: 20 years of research},
  author  = {Keith Rayner},
  journal = {Psychol. Bull.},
  year    = {1998},
  volume  = {124},
  number  = {3},
  pages   = {372--422},
  doi     = {10.1037/0033-2909.124.3.372},
  url     = {https://doi.org/10.1037/0033-2909.124.3.372},
  pmid    = {9849112}
}

@article{saeb2017approximate,
  author  = {Saeb, Sohrab and Lonini, Luca and Jayaraman, Arun and Mohr, David C. and Kording, Konrad P.},
  title   = {The need to approximate the use-case in clinical machine learning},
  journal = {GigaScience},
  volume  = {6},
  number  = {5},
  pages   = {gix019},
  year    = {2017},
  doi     = {10.1093/gigascience/gix019},
  url     = {https://doi.org/10.1093/gigascience/gix019}
}

@inproceedings{singh2023engagenet,
  title     = {Do {I} Have Your Attention: A Large Scale Engagement Prediction Dataset and Baselines},
  author    = {Singh, Monisha and Hoque, Ximi and Zeng, Donghuo and Wang, Yanan and Ikeda, Kazushi and Dhall, Abhinav},
  booktitle = {Proc. 25th ACM Int. Conf. Multimodal Interact. (ICMI)},
  pages     = {174--182},
  year      = {2023},
  publisher = {ACM},
  address   = {New York, NY, USA},
  doi       = {10.1145/3577190.3614164},
  url       = {https://doi.org/10.1145/3577190.3614164}
}

@article{wan2022hmdbiopad,
  title     = {A Wearable Head Mounted Display Bio-Signals Pad System for Emotion Recognition},
  author    = {Wan, Chunting and Chen, Dongyi and Huang, Zhiqi and Luo, Xi},
  journal   = {Sensors},
  volume    = {22},
  number    = {1},
  pages     = {142},
  year      = {2022},
  doi       = {10.3390/s22010142},
  url       = {https://doi.org/10.3390/s22010142},
  publisher = {MDPI}
}

@article{widmann2015digital,
  author  = {Widmann, Andreas and Schr{\"o}ger, Erich and Maess, Burkhard},
  title   = {Digital filter design for electrophysiological data -- a practical approach},
  journal = {J. Neurosci. Methods},
  volume  = {250},
  pages   = {34--46},
  year    = {2015},
  doi     = {10.1016/j.jneumeth.2014.08.002},
  url     = {https://doi.org/10.1016/j.jneumeth.2014.08.002}
}

@inproceedings{xgboost2016,
  author    = {Chen, Tianqi and Guestrin, Carlos},
  title     = {{XGBoost}: {A} Scalable Tree Boosting System},
  booktitle = {Proc. 22nd {ACM} {SIGKDD} Int. Conf. Knowl. Discovery Data Mining},
  year      = {2016},
  pages     = {785--794},
  publisher = {Association for Computing Machinery},
  address   = {New York, NY, USA},
  doi       = {10.1145/2939672.2939785},
  url       = {https://doi.org/10.1145/2939672.2939785}
}

@article{xiao2023multimodal,
  author  = {Xiao, Jun and Jiang, Zhujun and Wang, Lamei and Yu, Tianzhen},
  title   = {What Can Multimodal Data Tell Us about Online Synchronous Training:
             Learning Outcomes and Engagement of In-Service Teachers},
  journal = {Front. Psychol.},
  volume  = {13},
  pages   = {1092848},
  year    = {2023},
  doi     = {10.3389/fpsyg.2022.1092848},
  url     = {https://doi.org/10.3389/fpsyg.2022.1092848}
}

@article{zheng2017multimodal,
  author  = {Zheng, Wei-Long and Lu, Bao-Liang},
  title   = {A Multimodal Approach to Estimating Vigilance Using {EEG} and Forehead {EOG}},
  journal = {J. Neural Eng.},
  volume  = {14},
  number  = {2},
  pages   = {026017},
  year    = {2017},
  doi     = {10.1088/1741-2552/aa5a98},
  url     = {https://doi.org/10.1088/1741-2552/aa5a98}
}

@article{welch1967fft,
  author  = {Peter D. Welch},
  title   = {The Use of Fast Fourier Transform for the Estimation of
             Power Spectra: A Method Based on Time Averaging over Short,
             Modified Periodograms},
  journal = {IEEE Transactions on Audio and Electroacoustics},
  volume  = {15},
  number  = {2},
  pages   = {70--73},
  year    = {1967},
  doi     = {10.1109/TAU.1967.1161901}
}

@article{henrie2015measuring,
  author  = {Henrie, Curtis R. and Halverson, Lisa R. and Graham, Charles R.},
  title   = {Measuring Student Engagement in Technology-Mediated Learning: A Review},
  journal = {Computers \& Education},
  volume  = {90},
  pages   = {36--53},
  year    = {2015},
  doi     = {10.1016/j.compedu.2015.09.005},
  url     = {https://doi.org/10.1016/j.compedu.2015.09.005}
}

@article{whitehill2014faces,
  author  = {Whitehill, Jacob and Serpell, Zewelanji and Lin, Yi-Ching and
             Foster, Aysha and Movellan, Javier R.},
  title   = {The Faces of Engagement: Automatic Recognition of Student
             Engagement from Facial Expressions},
  journal = IEEE_J_AFFC,
  volume  = {5},
  number  = {1},
  pages   = {86--98},
  year    = {2014},
  doi     = {10.1109/TAFFC.2014.2316163},
  url     = {https://doi.org/10.1109/TAFFC.2014.2316163}
}

@article{karimah2022automatic,
  author  = {Karimah, Shofiyati Nur and Hasegawa, Shinobu},
  title   = {Automatic Engagement Estimation in Smart Education/Learning
             Settings: A Systematic Review of Engagement Definitions,
             Datasets, and Methods},
  journal = {Smart Learning Environments},
  volume  = {9},
  number  = {1},
  pages   = {31},
  year    = {2022},
  doi     = {10.1186/s40561-022-00212-y},
  url     = {https://doi.org/10.1186/s40561-022-00212-y}
}

\end{document}